\documentclass[aps,twocolumn,showpacs]{revtex4}
\usepackage{graphicx}
\begin{document}
\title{Multi-particle effects in non-equilibrium electron
tunnelling and field emission}
\author{Andrei Komnik and Alexander O. Gogolin}
\affiliation{Department of Mathematics, Imperial College, 180 Queen's Gate,
London SW7 2BZ, United Kingdom}
\date{\today}
\begin{abstract}
We investigate energy resolved electric current from various
correlated host materials under out-of-equilibrium conditions. We
find that, due to a combined effect of electron-electron
interactions, non-equilibrium and multi-particle tunnelling, the
energy resolved current is finite even above the Fermi edge of
the host material. In most cases, the current density possesses a
singularity at the Fermi level revealing novel manifestations of
correlation effects in electron tunnelling. By means of the
Keldysh non-equilibrium technique, the current density is
calculated for one-dimensional interacting electron systems and
for two-dimensional systems, both in the pure limit and in the
presence of disorder. We then specialise to the field emission
and provide a comprehensive theoretical study of this effect in
carbon nanotubes.
\end{abstract}
\pacs{03.65.X, 71.10.P, 73.63.F}

\maketitle
\section{Introduction}
At the early stages of the development of the quantum theory it
became clear that electron tunnelling processes are of
fundamental importance for condensed matter physics \cite{mahan}.
Electron tunnelling became an essential concept in such fields as
semiconductor physics, particle transport in mesoscopic physics,
field emission, and many others. In a multitude of setups,
tunnelling processes have attracted attention over decades
and continue to do so.

For the purposes of this work, we shall visualise tunnelling
events as taking place between a metallic host material and a
lead material (another metal, a semiconductor, or indeed vacuum
as in the field emission effect). The host and the lead are
separated from each other by a substantial potential barrier. The
lead is supposed to contain a detector measuring the current
density. The presence of a finite current immediately implies that
one is dealing with a non-equilibrium (steady-state) system.
Apart from the total current, the energy resolved current
$j(\omega)$, i.~e. the amount of current in the energy window
$\omega$ to $\omega+d\omega$, is an important characteristics of
the tunnelling process. The total current $J$ is then the
integral of $j(\omega)$ over all energies. In the field emission
(FE) setup, the energy resolved current have been measured for
different emitters over some 30 years (see \cite{plummer}, and
also references below). We shall therefore adopt the terminology
of the Gadzuk and Plummer's review article, \cite{plummer}, and
refer to $j(\omega)$ as the total energy distribution or TED. We
are not aware of any direct measurements of TEDs in tunnelling
contacts.

From the physical point of view, the most interesting aspect of
the TED is the part of the spectrum above the Fermi edge,
$\omega>E_F$. At the leading order in the tunnelling amplitude,
$j(\omega)$ is simply proportional to tunnelling probability
times the electron energy distribution function in the host,
$n(\omega)$. The latter is identically zero (at zero temperature)
above the Fermi edge due to the Luttinger theorem
\cite{luttingertheorem}. The high-energy tail is due to a
combined effect of non-equilibrium, multi-particle tunnelling
processes, and mutual electron-electron interactions in the host
material (for non-equilibrium multi-particle effects {\it per se}
are not sufficient to smear the Fermi surface \cite{caroli}). So,
we shall also refer to this high-energy tail of the energy
resolved current distribution as `secondary' current, `primary'
current being the one below the Fermi edge (i.~e. the one which is
proportional to the tunnelling probability). In the grand
picture, the secondary current is akin to the Auger effect though
it is, of course, a more complex phenomenon taking place not in
an atom but in a fully interacting metallic host. Measurements of
the secondary current (the trivial thermal broadening of the
primary current being subtracted off) can thus provide a valuable
source of information about the electronic correlations in the
host \cite{plummer,my}. The effect was indeed first discovered
experimentally -- in FE measurements by Lea and Gomer
\cite{1stexp}. Theoretical analysis of this phenomenon was done
by Gadzuk and Plummer soon afterwards \cite{gadzuk}. Following
the pioneering paper by Fowler and Nordheim \cite{fowler} these
authors used the connection between the FE problem and the
tunnelling problem and studied Boltzmann-like equations for the
particle-hole balance in the low density approximation. To remove
such restrictions and to put the theory on the modern footing, so
that it becomes applicable to strongly-correlated emitters, we
embark in this paper on investigating the issue by employing
Keldysh diagram technique, appropriate for this non-equilibrium
situation \cite{keldysh,LLX}. This method allows us to
consistently write down series in the tunnelling amplitude for
all quantities of interest and for the TED in particular.

Our motivation is in fact two-fold. On one hand, we think that
the secondary current phenomenon has not received sufficient
attention of theorists. The physics of the interplay between
non-equilibrium multi-particle tunnelling and electron
correlations is worth a deeper study. In particular, we
investigate in this paper what aspects of the electron
correlations are responsible for the current above the Fermi edge
and in what setups, other than the FE, such current can occur. On
the other hand, since the original work \cite{1stexp,gadzuk}
there have been considerable advances in the emitter technology.
Perhaps the most important recent development is the usage of
carbon nanotubes as field emitters. Carbon nanotubes display a
remarkable array of electronic and mechanical properties and are
of potential technological importance \cite{dekker}. Electron
transport in these systems has been thoroughly investigated.
While single-wall nanotubes (SWNTs) exhibit one-dimensional (1D)
Luttinger liquid (LL) type transport properties and are pretty
well understood, the theory \cite{sammlung1,sammlung2,sammlung3}
and experiment being in agreement \cite{tans1,bockrath1},
multi-wall nanotubes (MWNTs) are more complex systems subject to
intensive current debate. MWNTs are composed from many (at least
ten) concentric graphite shells. Current experiments on them are
consistent with two-dimensional (2D) diffusion with
characteristic weak localisation \cite{bachtold1} features and
zero-bias anomalies
\cite{schoenenberger,bachtold2,eg2001,mishchenko}. What concerns
us here is that, apart from other uses, carbon nanotubes are
expected to act as field emitters in high-resolution displays and
cathode tubes \cite{wong1,saito1}. While there have been several
experimental investigations  of the FE from carbon nanotubes (see
\cite{french1} and the main text for more references), both from
SWNTs and MWNTs, the relevant theory has been lacking. So, the
second leg of our motivation is to discuss existing FE
experiments on carbon nanotubes and make further theoretical
predictions for these systems.

Having in mind applications to carbon nanotubes and taking into
account the fact that FE usually occurs from a tip of the tube
(both for SWNTs and MWNTs) we narrow the following considerations
from a number of imaginable setups to an appropriate tip geometry.
Apart from this restriction, we intent to advance in this paper a
general discussion of non-equilibrium multi-particle tunnelling
from strongly-correlated 1D and 2D hosts paying special attention
to carbon nanotubes. Some of the results on 1D emitters
(applicable to SWNTs) have been announced in our recent letter
\cite{my}.

The paper is organised as follows. In the next Section we present
some qualitative considerations concerning the physical nature of
the TED. Section \ref{general} contains general
(model-independent) results. We identify the relevant Keldysh
diagrams contributing to the TED above the Fermi edge and then
perform a spectral analysis of the involved correlation functions.
A simple application of the developed theory is contained in
Section \ref{seclocalinteraction}, where we treat an electron
system with local interactions confined to the vicinity of the
tunnelling point. In Section \ref{luttinger} we analyse the TED of
particles tunnelling from LLs. In subsequent Sections \ref{pure}
and \ref{disorder}, the same problem is studied for correlated 2D
electron systems, respectively in the pure limit and in the
presence of a disorder potential. In Section \ref{FE} we
specialise to the case of field emission from carbon nanotubes.
Finally, in Section \ref{LS} we discuss a more sophisticated
two-stage tunnelling via a localised state. Summary and
conclusions Section completes the paper.

\section{Physical picture} \label{secII}
\begin{figure}
\includegraphics[scale=0.3]{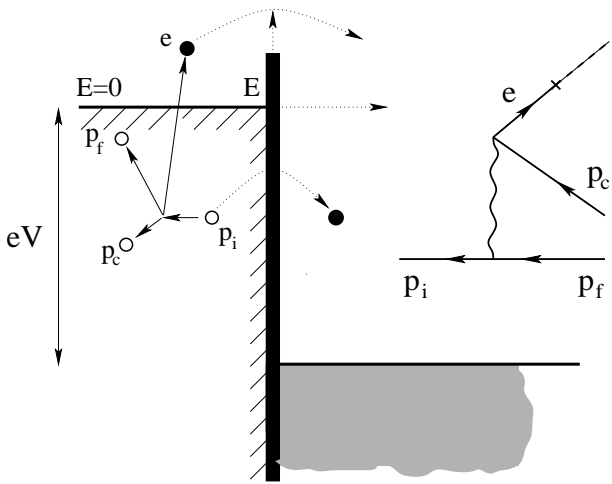}
\caption[]{\label{fig1}
Tunnelling junction biased by finite voltage $V$. The left electrode is the
interacting one while the right one is uncorrelated. Tunnelling electrons with
energies higher than $E_F$ are due to hot hole (denoted by
$p_i$) scattering with creation of electron-hole pairs $e-p_c$ (see
inset).}
\end{figure}

Before proceeding with calculations, let us elaborate on
qualitative origins of the TED high-energy tails.
The simplest setting to start with involves two noninteracting electrodes
with a tunnelling contact between them, see Fig.~\ref{fig1}.
Applying a finite voltage leads to a nonzero current through the
junction. Neglecting the charging effects and (for now) under the
assumption of constant electron densities of states and
transmission coefficient, the current is proportional to the
applied voltage $V$. On the right side of the contact, where the
chemical potential is supposed to be lower than on the left one,
the current is carried by the tunnelled electrons. On the left
side of the contact the current is carried by the holes moving in
the opposite direction away from the contact. As long as the
system is noninteracting, the TED of the electrons that tunnelled
out is uniform in the window between $E_F$ and $E_F-V$ and is
zero outside (this is evident but can be confirmed by simple
calculation in the spirit of \cite{caroli}, which we omit). From
now on we set $e=1$. $E_F$ is the Fermi level of the left
electrode.

This picture changes drastically if interactions between the
electrons are switched on. We restrict our considerations to the
case when only the left contact -- the host -- is correlated (see
also discussion in the next Section). Then the difference between
the actual energy distribution function in the lead and the
noninteracting distribution function gives the TED of particles
tunnelling out of the host. The holes left behind by electrons
tunnelling out below the Fermi energy (`primary' electrons)
experience scattering from the electron sea, thereby creating
electron-hole pairs. Contrary to the problem of hot electron
relaxation (see e.~g.~\cite{meden}, and references therein), we
are dealing here with a \emph{flow} of hot holes in a steady
state. Also, we are interested in a different object: the energy
distribution functions rather than momentum distribution
functions (the latter do, of course, have a non-zero tail above
the Fermi momentum solely due to correlations, without a need to
include tunnelling \cite{agd}). The emerging `secondary'
electrons can also be carried over to the lead by a successive
tunnelling process, as shown in Fig.~\ref{fig1}. This phenomenon
can be regarded as condensed matter analogy of the Auger process
known from the atomic physics. Since the energy of the holes,
measured from the Fermi edge $E_F$, can not exceed $V$, the upper
limit for the energy the secondary electrons can acquire in the
electron-hole pair creation process is given by $V$ as well.
At higher orders in the tunnelling probability this threshold is increased
so that the secondary electrons can, in fact, become more energetic than $V$.
In this case they ought to be successively scattered either from
another hot hole or from another hot electron.
This is, however, a process of higher order not only in the tunnelling but
also in the interaction.
Therefore secondary electrons with energies between $(n-1) V$ and $n V$
emerge in the processes of $2n$th order with respect to the tunnelling amplitude
as well as the interaction constant.
Obviously, in equilibrium $V=0$ all the high energy tails vanish, thereby
restoring the Fermi edge in accordance with the Luttinger theorem.
So, non-trivial interaction effects are encoded in the high-energy behaviour
of the non-equilibrium TED.

Theoretically, the most interesting limiting case is the
behaviour of the TED just above the Fermi edge, when certain
universality can be expected. Out of the considerations of
Ref.~\cite{gadzuk} a divergent TED emerged, with (simplifying
matters) the singularity approximately of the form $j(\omega)\sim
1/(\omega-E_F)$. This turned out to be roughly consistent with
measurements at the time \cite{1stexp}. As detailed in the
following Sections, the present study supports the view that the
limiting form of the TED strongly depends on the nature of the
host material and the geometry of the setup (within a given
material type though, some universality does set in, so for
point-contacted LLs we find a power-law behaviour, etc.).

\section{Statement of the problem and general results} \label{general}
We now formalise the problem by writing down
the relevant tunnelling Hamiltonian:
\begin{equation} \label{ham0}
 H = H[c] + H[\psi] + \gamma \left[ \psi^\dag(0) c(0) + c^\dag(0) \psi(0)
 \right] \; .
\end{equation}
Here $\gamma$ is the tunnelling amplitude and $c$ and $\psi$ are the
annihilation operators for the electrons
in the lead and in the host, respectively.
The unperturbed part of the Hamiltonian $H_0=H[c] + H[\psi]$ describes
two decoupled electron systems at different chemical potentials
$\mu_\psi-\mu_c=V>0$.

A clarification on the following points is in order.

{\bf (i)} As we only want to consider a tip geometry, we have
explicitly assumed that the tunnelling occurs only locally at the
location of the tip: $x=0$. In reality, of course, there is a
small area over which the tunnelling takes place (strictly speaking,
a real-space integral is required in the tunnelling term in
Eq.(\ref{ham0})). As this has no qualitative influence, we shall
keep writing simple as long as we can. The locality assumption is
natural for carbon nanotubes because of the very shape of these
objects. It is however also justified for most bulk interfaces
and field emitters where, because of the roughness of the surface
and the pronounced exponential dependence of the penetration
coefficient on the distance between the electrodes, the main
contribution to the current comes from only few points between
the electrodes.

{\bf (ii)} Related to the above is the question of the energy
(momentum) dependence of the tunnelling amplitude. In real
systems it is energy dependent. So, for the FE setup the relevant
energy scale is determined by a combination of the work function
and the applied field. It is important to keep in mind, however,
that as the tunnelling amplitude is a single electron property,
regarding its energy dependence there is no special significance
to the Fermi edge. Therefore, when addressing observables
determined by scattering processes taking place close to the
Fermi surface, it is quite safe to neglect the energy dependence
of the tunnelling amplitude and replace it by a constant,
$\gamma$ ($\gamma^2$ being proportional to the transmission
coefficient of the potential barrier at the Fermi energy: ${\cal
D}(E_F)$.) In the literature on the subject the tunnelling term is
often quoted in the momentum representation in the form
$\gamma_{pk}\psi^\dagger_p c_k$ plus conjugate with some
unspecified matrix elements $\gamma_{pk}$. We take the view that,
for calculating quantities related to the Fermi surface, this
would only complicate formulae without conceptual gain. There
will be, however, instances in the following when the energy
dependence of the tunnelling amplitude is important, like when
there is a localised state or when we discuss generalisations of
the Fowler--Nordheim relations for nanotubes. In these cases
(Sections \ref{FE} and \ref{LS}) we shall take the relevant
energy dependence fully into account.

{\bf (iii)} Throughout this paper we assume that the lead is
non-interacting. Indeed one could not otherwise separate the
effect of correlations occurring in the host from those occurring
in the lead, which would seriously hamper meaningful
interpretation of measurements on such systems. This assumption is
justified for the FE setup, save for the Boersh effect, which is
not believed to be important for carbon nanotubes (see
\cite{fransen} and \cite{my}). For a general tunnelling junction
setup it would not be correct to {\it a priori} neglect
correlations in the lead. However, with care it is possible to
realise a reasonable setup involving, for instance, a nanotube in
contact with a low carrier density semiconductor or a quantum wire
opening up into a higher dimensional lead preferably screened by
a nearby gate (in fact most conductance measurements on quantum
wires are nowadays interpreted in terms of a junction with a
non-interacting lead \cite{tarucha,glazman}).

In the bulk of this paper we shall be investigating the secondary
current $j(\omega>0)$ (we set $E_F=0$ from now on). It is a
plausible statement that the secondary current should be
proportional to the high-energy tail $n(\omega>0)$ of the
electron energy distribution function {\sl inside} the emitter tip
(calculated to all orders in the tunnelling amplitude). Indeed
the latter quantity corresponds to the amount of electrons
available for tunnelling at a given energy $\omega$, while the
proportionality factor contains (unessential) information about
how these electrons are then carried over to the detector. In the
first part of this Section we establish this statement.

\begin{figure}
\includegraphics[scale=0.3]{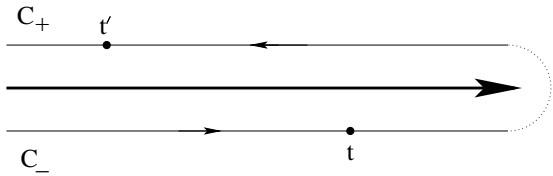}
\caption[]{\label{contour} Schematic representation of the Keldysh contour.
The lower branch is time-ordered, while the upper
branch is anti-time-ordered.
}
\end{figure}

Since we are dealing with non-equilibrium phenomena we resort to the
Keldysh formalism. We denote by  $G(x,x';t-t')$ the
generalised Green's function for electrons in the lead,
\begin{equation} \label{first}
G(x,x';t-t') = -i \langle T_C [c(x,t)
c^\dag(x',t')\,S_C] \rangle_0 \, ,
\end{equation}
the Green's function in the host, $g$, being defined by an analogous
formula.
Here $T_C$ stands for the ordering operation along the Keldysh
contour, shown in Fig.~\ref{contour}, and $x$ corresponds to a
set of coordinates specifying the electron states in the lead.
(It should be understood as a distance from the emitter tip plus
possibly a transverse channel index which we suppress as it plays
no important part in the following). The average in
Eq.~(\ref{first}) is taken over the ground-states of the
unperturbed Hamiltonian $H_0$ and the contour $S_C$-matrix is
responsible for tunnelling events:
\begin{eqnarray} \label{Smatrix} \nonumber
S_C = T_C \exp\left(-i \gamma \int\limits_C \, dt \, [ \psi^\dag(t) c(t) +
c^\dag(t) \psi(t) ]\right) \, .
\end{eqnarray}
Note that because the system is not translationally invariant, all
Green's functions depend on both coordinates. However, all
functions still only depend on the time differences because the
system is in steady state. The time integration in
Eq.~(\ref{first}) is along the contour $C$. Keldysh
disentanglement of the time variables results in expressions with
integrations only along the real axis. Then, four different
Green's functions emerge in accordance with four possibilities to
arrange the times $t$ and $t'$ along the contour. This placement
of the time variables is reflected in additional superscripts of
the Green's functions. For instance, if the time $t$ lies on the
time-ordered part of the contour and $t'$ on the
anti-time-ordered one, see Fig.~\ref{contour}, evaluation of
Eq.~(\ref{first}) at the zeroth order in the tunnelling (indicated
by a subscript $0$) yields:
\begin{eqnarray}
G^{-+}_0(x,x';t-t') &=& -i \langle T_C [c(x,t)
c^\dag(x',t') ]\rangle_0 \nonumber \\ \nonumber &=& i \langle
c^\dag(x',t') c(x,t) \rangle_0 \, .
\end{eqnarray}
Its counterpart with an interchanged orientation
of time variables ($t$ on the $C_+$ and $t'$ on the $C_-$) is
\begin{eqnarray} \nonumber
G^{+-}_0(x,x';t-t') = -i \langle c(x,t) c^\dag(x',t') \rangle_0 \, .
\end{eqnarray}
Green's function in which both time variables lie on the same side
of the contour are the usual time-ordered and anti-time-ordered ones:
\begin{eqnarray}
G^{--}_0(x,x';t-t') &=& -i \langle T [c(x,t) c^\dag(x',t')
]\rangle_0 \, , \nonumber \\ \nonumber G^{++}_0(x,x';t-t') &=& -i
\langle \widetilde{T}[ c(x,t) c^\dag(x',t') ]\rangle_0 \, ,
\end{eqnarray}
where $\widetilde{T}$ denotes the anti-time-ordering operation.

\begin{figure}
\includegraphics[scale=0.3]{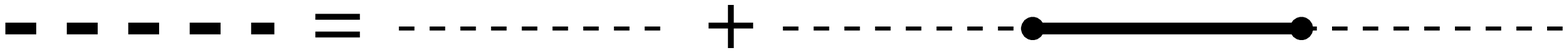}
\caption[]{\label{equation} Schematic representation of the equation
(\ref{impform}). Solid line denotes the electron's Green's function of the
lead and the dashed ones stand for particles in the host.
Thick lines they correspond to exact Green's functions to all
orders in tunnelling and interactions while thin lines represent Green's
functions with tunnelling neglected.
}
\end{figure}

The local electron energy distribution function $N(x,\omega)$ in
the lead, which we shall also call TED by abuse of terminology,
is given by the defining relation with one of the Keldysh Green's
functions (see e.~g.~\cite{LLX}):
\begin{equation} \label{defEDF}
 N(x,\omega) = - i G^{-+}(x,x; \omega) \; .
\end{equation}
The lead being non-interacting, the TED of the tunnelling
particles is, up to a pre-factor, given by above
Green's function after subtracting off
the equilibrium distribution function.
Indeed, by examining the perturbative expansion in
the tunnelling amplitude $\gamma$ and keeping in mind
the fact that there is no correlations in the lead,
one can easily establish the following important identity,
\begin{eqnarray} \label{impform}
&~&~G(x,x'; t-t') = G_0(x,x';t-t') \\
\nonumber &~&~+\gamma^2 \int\limits_C\int\limits_C dt''dt'''
G_0(x,0;t-t'') g(0,0;t''-t''') \\
\nonumber &~&~\times G_0(0,x';t'''-t') \; .
\end{eqnarray}
The corresponding diagram is shown in Fig.~\ref{equation}.
The function $g(0,0;t-t')$ appearing in this relation is the
exact one, so the formula is valid for arbitrary
interactions in the host and to all orders in the tunnelling amplitude.

By disentangling the Keldysh indices and changing over to the
energy representation, we extract the Green's function of interest
from the general expression, (\ref{impform}),
\begin{eqnarray} \label{dis}
&& G^{-+}(x,x'; \omega) = G^{-+}_0(x,x';\omega) \\
\nonumber &+& \gamma^2  \sum_{i,j=\pm} (ij)
\, G^{-i}_0(x,0;\omega) g^{ij}(0,0;\omega) G^{j+}_0(0,x';\omega)\, .
\end{eqnarray}
Plugging this expression into Eq.~(\ref{defEDF})
one obtains the complete TED at all energies.
Let us take a closer look at this relation.
All terms on the right-hand-side, apart of the one with
$(i,j)=(-,+)$, contain the unperturbed $G_0^{-+}$ function.
The latter is, in turn, proportional to the TED in equilibrium.
Therefore, no matter what the Green's functions of the interacting
fermions look like, all these contributions represent the TED below the
Fermi energy (where the first term is the dominant contribution).
The term in Eq.~(\ref{dis}) with $(i,j)=(-,+)$ is not constrained in such
way and can, in principle, contribute above the Fermi edge.
Therefore the high-energy part of the TED is given by
\begin{eqnarray} \label{EDFexp}
 N_{>}(x;\omega) &=& -i\Theta(\omega) G^{-+}(x,x; \omega) \\ \nonumber &=&
 - \gamma^2 G^{--}_0(x,0;\omega) g_>^{-+}(0,0;\omega) G^{++}_0(0,x;\omega)
\; ,
\end{eqnarray}
where $\Theta(\omega)$ is the Heaviside step function and
$g_>^{-+}(0,0;\omega)$ stands for the high energy part of the corresponding
Green's function.

Similar properties can be established for the energy resolved
current $j(\omega)$, when we have
\begin{eqnarray} \nonumber
 j(\omega) = \frac{1}{\pi} \int \, dk \, v_k \,  N(k;\omega) \,
\end{eqnarray}
where $v_k$ is the velocity of the particle with wave number $k$.
Making use of Eqs.~(\ref{defEDF}) and (\ref{dis}) for the
high-energy part of the current one obtains,
\begin{eqnarray} \label{tokexp}
j_>(\omega) &=& -i \frac{\gamma^2}{\pi} \, g^{-+}_>(0,0;\omega) \\ \nonumber
&\times& \int \, dk \, k \int \, d(x-x') \, e^{i k (x-x')}\\
\nonumber &\times& \left[G^{--}_0(x,0;\omega)
G^{++}_0(0,x';\omega) \right] \, ,
\end{eqnarray}
where the product in brackets is actually translationally
invariant (only depends on $x-x'$), reflecting the fact that the
excess particles only travel in one direction, away from the
contact, in the (non-interacting) lead.

We now observe that both Eq.~(\ref{EDFexp}) and Eq.~(\ref{tokexp})
are proportional to the high-energy part of the TED of particles
at the tip of the interacting host,
\begin{eqnarray} \label{tokexp1}
n_>(\omega) = -i g^{-+}_>(0,0;\omega) \, ,
\end{eqnarray}
thereby proving the statement put forward at the beginning of
this Section. We stress again that $n_>(\omega)$ is supposed to
be exact both with respect to the interaction and the tunnelling.
Indeed, as one can easily see, this quantity is zero for a
noninteracting system, even if tunnelling is taken into account
to all orders. It is still zero for interacting hosts if
tunnelling is neglected or if the system is in equilibrium
(Luttinger theorem). Situation changes dramatically in the case
of an interacting host, finite tunnelling and finite bias voltage.
In the rest of this Section we shall explore what statements can
be made about the TED without assuming a concrete model for the
host material (other than that there is a two-body interaction
present).

\begin{figure}
\includegraphics[scale=0.3]{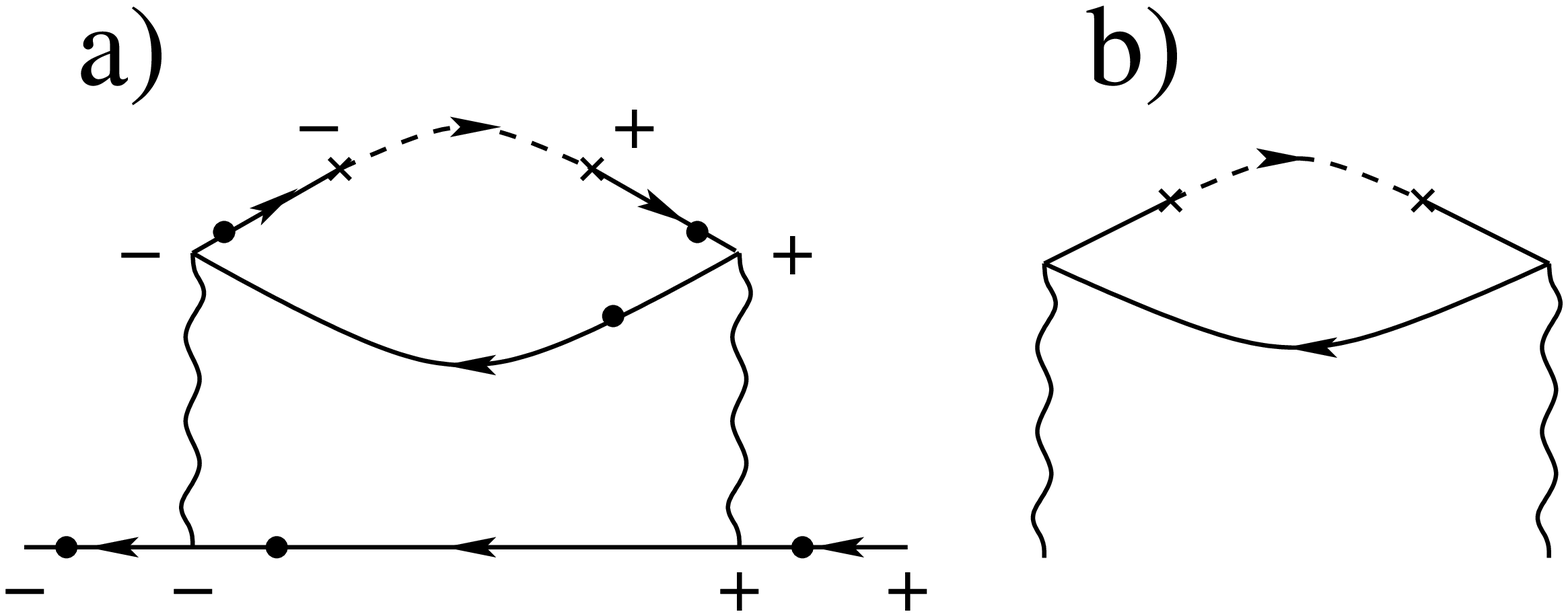}
\caption[]{\label{fig3} a) The only second order diagram
contributing to the TED above the Fermi energy. Solid lines
correspond to host electrons and the dashed one to those in the
lead while crosses denote the tunnelling vertices. Wiggly lines
represent interactions. b) This diagram has to be inserted into
a) at points denoted by circles in order to obtain high-order
diagram for the TED. }
\end{figure}

As mentioned in the Introduction, there is a scattering process
that allows for creation of electrons above the Fermi energy. Let
us start with perturbative analysis. At the second order in the
interaction, the corresponding diagram for the Green's function
can be constructed by annihilating the particles in the same way
they were created. This essentially converts the scattering
amplitude for a given process into the corresponding probability.
The result is shown in Fig.~\ref{fig3}~a). Now we need to establish
the way the vertices can be decorated with the Keldysh
indices. Trivially the outmost points should have index $-$ on
the outward leg and $+$ on the inward leg in accordance with the
type of function ($-+$) we are calculating. Furthermore, there is
only one possibility to assign indices to interaction vertices.
In order to obtain a contribution above the Fermi energy the
inward Green's function should be of $++$ or anti-time ordered
type while the outward one has to carry $--$ indices, because all
other possibilities contain at least one $G^{-+}_0$ factor which
forces the diagram to vanish above $E_F$.
We are left with two cross vertices of the inserted Green's function.
Since we do not have \emph{a priori} knowledge about them,
we have checked all possibilities explicitly.
It turns out that the only possibility is to insert the function
$G^{+-}_0$ because in all other cases the diagram vanishes for
$\omega > 0$.
Hence the only second-order diagram contributing to the TED
above the Fermi energy is the one shown in Fig.~\ref{fig3}~a).

Let us now consider all orders in the interaction but remain at
the second order in the tunnelling amplitude thereby also
formally justifying the above choice of the Keldysh indices.
The way to proceed is to analyse the relevant Green's functions
in the Lehmann-type spectral representation.
In the time domain the local (tip) Green's function of
interest $g^{-+}(t_1,t_2)$ is given by
\begin{eqnarray} \nonumber
g^{-+}(t_1,t_2) = i \langle \psi^\dag(t_2) \psi(t_1)
\rangle = -i \langle T_C [\psi(t_1^-)\psi^\dag(t_2^+)] \rangle \; ,
\end{eqnarray}
where the superscripts indicate that the time variables $t_1$ and $t_2$
lie on the time-ordered ($T$) and anti-time-ordered ($\widetilde{T}$) parts of
the Keldysh contour, respectively.
Performing the straightforward $S$-matrix expansion in powers of the tunnelling
amplitude $\gamma$ one obtains at the lowest non-vanishing order
(second order in $\gamma$):
\begin{eqnarray} \label{gsum}
\delta g^{-+}(t_2,t_1) &=& \gamma^2 \int \, dt_3 \, dt_4 \, \sum_{ij = \pm}
{\cal K}^{-+}_{ij}(t_2,t_1; t_3,t_4) \nonumber \\
&\times& G^{ij}(t_3,t_4) \, ,
\end{eqnarray}
where ${\cal K}^{-+}_{ij}(t_2,t_1; t_3,t_4)$ are the following four-point
correlation functions:
\begin{eqnarray} \label{gammadef}
 {\cal K}^{-+}_{+-}(t_1,t_2; t_3,t_4) &=& \langle \widetilde{T}
[\psi^\dag(t_2)\psi^\dag(t_3)] T[\psi(t_1) \psi(t_4)] \rangle_0
\, , \nonumber \\
 {\cal K}^{-+}_{-+}(t_1,t_2; t_3,t_4) &=& \langle \widetilde{T}
[ \psi(t_4) \psi^\dag(t_2)] T [ \psi(t_1) \psi^\dag(t_3)] \rangle_0
\, , \nonumber \\
 {\cal K}^{-+}_{--}(t_1,t_2; t_3,t_4) &=& \langle \psi^\dag(t_2)
T [\psi(t_1) \psi^\dag(t_3) \psi(t_4)] \rangle_0 \, , \nonumber \\
 {\cal K}^{-+}_{++}(t_1,t_2; t_3,t_4) &=& \langle \widetilde{T}
[ \psi^\dag(t_2) \psi^\dag(t_3) \psi^\dag(t_4)] \psi(t_1)
\rangle_0 \,
\end{eqnarray}

We shall first find the spectral representation of the term
containing ${\cal K}^{-+}_{+-}$,
\begin{eqnarray} \label{EDFB}
\delta^{(1)} g^{-+}(t_1,t_2) = \gamma^2 \int \,
\frac{d\epsilon}{2 \pi} G^{+-}(\epsilon) \nonumber \\
\times \int \, dt_3 dt_4 \, e^{-i \epsilon (t_3-t_4)}
{\cal K}^{-+}_{+-}(t_1,t_2; t_3,t_4) \nonumber \\
= -i \gamma^2 \int_{-V}^{\infty} \, d\epsilon \,
\rho_c(\epsilon) P^{-+}_{+-}(t_1,t_2;\epsilon) \, ,
\end{eqnarray}
where we have used the
bare $-+$ Green's function in the lead,
\begin{eqnarray} \label{G0omega}
 G^{+-}_0(\epsilon) = -i 2 \pi \Theta(\epsilon + V) \rho_c(\epsilon) \, ,
\end{eqnarray}
containing the local density of states $\rho_c(\epsilon)$ in the lead.
Partial Fourier transform of the correlation function appearing in
the above formula is defined by
\begin{eqnarray} \nonumber
 P^{-+}_{+-}(t_1,t_2;\epsilon) = \int \, dt_3 \, dt_4 \,
e^{-i \epsilon (t_3-t_4)} {\cal K}^{-+}_{+-}(t_1,t_2; t_3,t_4) \, .
\end{eqnarray}
According to definition (\ref{gammadef}), this correlation function contains
two different time orderings. Writing them down explicitly and
inserting between every two $\psi$ operators a complete set of exact states,
it is possible to perform all time integrations.
This procedure is a straightforward generalisation of the standard one for
the equilibrium case \cite{agd}.
The result is
\begin{eqnarray} \label{pis}
P^{-+}_{+-}(t_1,t_2;\epsilon) = \sum_\mu e^{-i(E_\mu+\epsilon)(t_1-t_2)}
|{\cal B}^{(1)}_\mu (\epsilon)|^2 \, ,
\end{eqnarray}
with ${\cal B}^{(1)}_\mu (\epsilon)$ defined by
\begin{eqnarray} \nonumber
{\cal B}^{(1)}_\mu(\epsilon) = \sum_\nu a_{\mu \nu}a_{\nu
0}\left[\frac{1}{E_\mu-E_\nu+\epsilon+i0} +\frac{1}{E_\nu+\epsilon-i0}\right]
\; , \nonumber
\end{eqnarray}
where Greek indices count all possible excited states of the
system with energies $E_{\nu, \lambda, \mu}$ and $a_{\mu \nu}$
stand for matrix elements of the operator $\psi$ , $a_{\mu
\nu}=\langle \mu| \psi | \nu \rangle$. In order to obtain the
actual correction to the TED we plug the two last equations back
into Eq.~(\ref{EDFB}) and compute the Fourier transform of the
latter with respect to the time difference $t_1-t_2$,
\begin{eqnarray} \label{specres1}
\delta n^{(1)}(\omega)= - i \int \, d(t_1-t_2) \, e^{i \omega (t_1-t_2)}
\, \delta^{(1)} g^{-+}(t_1,t_2) \nonumber \\
 = - 2 \pi \gamma^2 \sum_\mu \Theta(V - E_\mu - \omega) |{\cal
 B}^{(1)}_\mu(-V)|^2 \, .
\nonumber
\end{eqnarray}
Obviously, all $E_\mu$'s are larger than the ground state energy $E_0$
(which we have set to zero).
Therefore the upper boundary for $\omega$ is given by $V$.

The remaining three terms in Eq.~(\ref{gsum}) can be treated in a
similar manner. Repeating steps leading to (\ref{pis}), we obtain
for the term containing the second four-point correlation
function in Eq.~(\ref{gammadef}):
\begin{eqnarray} \nonumber
 P^{-+}_{-+}(t_1,t_2;\epsilon) =
\sum_\mu e^{i (E_\mu-\epsilon)(t_1-t_2)} |
{\cal B}^{(2)}_\mu(\epsilon) |^2 \, ,
\end{eqnarray}
with
\begin{eqnarray} \nonumber
{\cal B}^{(2)}_\mu (\epsilon) = \sum_\nu \left[ \frac{a_{0\mu}
a^*_{\mu \nu}}{E_\mu - \epsilon + i0} + \frac{a^*_{0 \mu}
a_{\mu \nu}}{E_\nu - E_\mu - \epsilon - i0} \right] \, ,
\end{eqnarray}
the corresponding contribution to the TED being
\begin{eqnarray} \label{specres2}
 \delta^{(2)}n (\omega) = 2 \pi
\gamma^2 \sum_\mu \Theta(-V - E_\mu - \omega) |{\cal B}^{(2)}_\mu (V) |^2 \, .
\end{eqnarray}
Contrary to Eq.~(\ref{specres1}), this term is bounded by $-V$
from above and hence does not contribute to the high-energy part
of the TED. One can use an even simpler argument in order to
show that the last two four-point correlation functions in
(\ref{gammadef}) do not contribute above the Fermi edge either.
By inserting only one complete set of states between the
time-ordered operators (i.~e. at the break of the time-ordering)
one obtains (for simplicity we set $t_1=0$)
\begin{eqnarray} \nonumber
 P^{-+}_{--}(0,t_2;\epsilon) &=& \int \int \, dt_3 \,
dt_4 \, e^{-i \epsilon (t_3-t_4)} \\ &\times& \sum_\nu
a^*_{0 \nu} e^{-i E_\nu t_2} \langle \nu | T [\psi(0)
\psi^\dag(t_3) \psi(t_4) ]| 0 \rangle \,  \nonumber
\end{eqnarray}
for the third term.
The corresponding correction to the TED is then given by
\begin{eqnarray}
 &~& \delta^{(3)} n(\omega) = i \gamma^2
\sum_\nu a^*_{0 \nu } \delta(-E_\nu - \omega) \int d\epsilon \,
G^{--}_0(\epsilon) \nonumber \\ \nonumber
&\times& \int dt_3 d t_4
\langle \nu | T [\psi(0) \psi^\dag(t_3) \psi(t_4)] | 0 \rangle \,
.
\end{eqnarray}
Since all $E_\nu$'s are always positive this expression is nonzero
only for negative energies $\omega<0$.
The same is true for the last four-point correlation function
${\cal K}^{-+}_{++}$.

\begin{figure}
\includegraphics[scale=0.4]{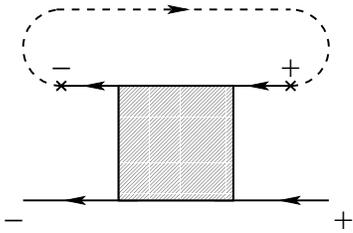}
\caption[]{\label{fig4}
Generic diagram giving contributing to the TED above the Fermi energy.
The four-point correlation function ${\cal K}^{-+}_{+-}(t_1,t_2;t_3,t_4)$ is
represented by the shaded square. }
\end{figure}

The above approach is quite general (i.~e. valid for all kinds of
interacting host materials) but it is also inconclusive in the sense
that the actual energy dependence of the TED is determined by
spectral weights encoded in the structure of matrix elements
$a_{\mu \nu}$ that is different for different hosts. One positive
result of the spectral method, however, is that we have
identified the Keldysh four-point correlation function
responsible for the secondary current effect:  ${\cal
K}^{-+}_{+-}$, see Fig.~\ref{fig4}. For some models we shall
calculate this correlation function exactly to all orders in the
interaction, for other models we'll resort to perturbative
expansions: note that diagram a) of Fig.~\ref{fig3} is a special
case (second-oder expansion) of the general Fig.~\ref{fig4} term.

Let us conclude the present Section by making one more observation
of a general character.
Returning back to the real time representation of the lead Green's
function (\ref{G0omega})
and doing simple manipulations
with the time integrations, we arrive at the alternative representation
for the TED:
\begin{eqnarray} \label{com}
&~&~n_>(\omega) = - 2i \gamma^2 \int\limits_{-\infty}^\infty  dt
\frac{e^{i (\omega-V)t}}{t+i \alpha} \int\limits_{-\infty}^\infty d\tau_1
\int\limits_{-\infty}^\infty d\tau_2e^{i\omega(\tau_1+\tau_2)} \nonumber  \\
&~&~\times
\langle \widetilde{T}[\psi(\tau_1)\psi(0)]
T[\psi(t+\tau_1)\psi(t+\tau_1+\tau_2)]\rangle_0 \, ,
\end{eqnarray}
where $\alpha$ stands for the short-time (high-energy) cutoff
inversely proportional to the conductance band-width $D$. For the
sake of simplicity, we have assumed a constant density of states
in the lead but this assumption is not crucial.
We now observe that if the $T$-ordering operation in the above
formula were dropped, the expression would vanish. Indeed,
inserting some complete sets one finds that then the integrand in
the above, proportional to
\begin{equation} \nonumber
\sim\langle 0 |\psi(t+\tau_1)|\nu\rangle\langle \nu|
\psi(t+\tau_1+\tau_2)|0\rangle
\sim e^{iE_\nu\tau_2} \, ,
\end{equation}
becomes an analytic function of the time variable $\tau_2$ in the
upper half plane and hence has vanishing Fourier transform for
$\omega>0$. By re-arranging time integrations in a different way,
one can easily show that the same statement is true about the
$\widetilde{T}$-ordering operation. Subtracting off the unordered
correlation function we arrive at the following remarkable
representation for the TED:
\begin{eqnarray} \label{combis}
&~&~n_>(\omega) = - 2i \gamma^2 \int\limits_{-\infty}^\infty  dt
\frac{e^{i (\omega-V)t}}{t+i \alpha} \int\limits_{0}^\infty d\tau_1
\int\limits_{0}^\infty d\tau_2e^{i\omega(\tau_1+\tau_2)} \nonumber  \\
&~&~\times
{\cal R}(\tau_1,0;t+\tau_1,t+\tau_1+\tau_2) \, ,
\end{eqnarray}
with
\begin{eqnarray} \label{Rdef}
{\cal R} (t_1,t_2;t_3,t_4)= \langle \{
\psi^\dag(t_1),\psi^\dag(t_2) \} \{ \psi(t_3),\psi(t_4) \}
\rangle \, ,
\end{eqnarray}
\vspace*{0.0cm}

\noindent where $\{.,.\}$ stands for the anti-commutator. The
advantage of this representation is that it is explicitly
vanishing in the non-interacting case as the field operators are
then anti-commuting at all times. The latter does not take place
in interacting systems. Generally, anti-commuting interacting
Fermi operators at different times would result in a complicated
object. However, there are systems for which the so-called
\emph{braiding relations} hold:
\begin{eqnarray} \label{braiding}
 \psi(t_1) \psi(t_2) = e^{i \varphi(t_1-t_2)} \psi(t_2) \psi(t_1) \,.
\end{eqnarray}
The exact shape of the phase function $\varphi(t)$ depends on the
system in question but it is usually proportional to the
interaction constant. The braiding relation reflects the time
evolution of an anti-commutators in particular models and can be
used to simplify (\ref{Rdef}) to
\begin{widetext}
\begin{eqnarray} \label{braiding2}
 {\cal R}(\tau_1,0 ; t+\tau_1,t+\tau_1+\tau_2)
 = (1+e^{-i \varphi(\tau_1)})(1+e^{-i \varphi(-\tau_2)})
\langle \psi^\dag(\tau_1)\psi^\dag(0)
\psi(t+\tau_1)\psi(t+\tau_1+\tau_2) \rangle \,
\end{eqnarray}
\end{widetext}
from which formula it is clear that the expansion of $n_>(\omega)$
in the interaction constant starts (at least) at the second order.
In the next Section we present two models where such braiding
relations exist.

In the following four Sections we apply the formalism developed
here to various physical systems.

\section{Models with local interactions} \label{seclocalinteraction}
We start with the simplest (toy) model we could think of:
the interactions are only present at one point - the tip of the
emitter.
The host Hamiltonian entering Eq.~(\ref{ham0}) is then of the form
\begin{equation}
H[\psi]=H_0[\psi]+ H_i[\psi]
\end{equation}
with the Hubbard on-site repulsion for the interaction term
\begin{equation} \label{intdefinition}
H_i = U_0 \psi^\dag_\uparrow(0) \psi^\dag_\downarrow(0) \psi_\downarrow(0)
\psi_\uparrow(0) \, .
\end{equation}
We take a half-infinite electron system with linear
dispersion relation for the left electrode
\begin{eqnarray} \nonumber
H_0[\psi]= -i \sum_s \int\limits_0^\infty \, dx \,
\Big[ \psi_{sR}^\dag(x) \partial_x \psi_{sR}(x) -
\psi_{sL}^\dag(x) \partial_x \psi_{sL}(x)\Big]  \; ,
\end{eqnarray}
and the same for the lead (but, of course, at chemical potential
of $-V$). Here $\psi_{sR(L)}$ stands for the annihilation operator
of the right(left)-moving electron species with spin $s$. To
simplify the formulae we have set $\hbar v_F=1$.

At first sight such model seems unphysical.
However, it can be regarded as a special case of a
correlated quantum dot contacted by two
noninteracting electrodes, see e.~g. Ref.~[\onlinecite{wingreen}], where the
contact to one of them is nearly perfect while the contact to the
other one is weak.
One should then decorate the electron field operator with a tunnelling
channel (transverse quantisation) index.
Its presence does not affect qualitative results, so we
drop the channel index in this Section.

For the half-infinite host, the bare Green's functions have
image structure of the form
\begin{equation} \label{goodformula}
g_{s0}(x,x';\omega) = g_0(x-x';\omega) - g_0(x+x';\omega) \; ,
\end{equation}
and are diagonal and $s\to -s$ symmetric in the spin space. The
Green's functions appearing on the right-hand-side in the above
formula are the translationally invariant (bulk) ones for the
corresponding spinless infinite system.
The simplest Green's functions are the retarded and advanced ones,
\begin{equation} \nonumber
 g_0^{R}(x;\omega) = [g_0^A(x;\omega)]^* = -i e^{i (k_F + \omega) |x| } \,
\end{equation}
to which the diagonal Keldysh functions are related via
\begin{equation} \nonumber
g_0^{--}(x;\omega) = \Theta(\omega) g_0^{R}(x;\omega)+
\Theta(-\omega) g_0^{A}(x;\omega)
\end{equation}
and $g_0^{++}(x;\omega)=-[g_0^{--}(x;\omega)]^*$,
while the off-diagonal Keldysh functions are:
\begin{eqnarray} \nonumber
g_0^{+-(-+)}(x;\omega)=i 2 \Theta(\pm \omega) \cos \Big[(k_F +
\omega)x\Big] \; .
\end{eqnarray}

Plugging these expressions into Eq.~(\ref{goodformula}) one
obtains all Keldysh components for the half-infinite system.
However, using them directly to evaluate $g^{-+}_{s>}(0,0;\omega)$
for electrons with spin orientation $s$ causes a technical problem
as at the origin all wave functions of the half-infinite system
vanish and so do the correlation functions. The reason for that
is the open boundary condition imposed on the wave functions.
Therefore we are in need of a regularisation. We now assume that
the underlying lattice model has a lattice constant $a_0$ so that
the tunnelling occurs between points with the spatial coordinate
$x<a_0$ in the left subsystem as well as in the right one.
Obviously, such regularisation does not influence the physics but
only creates some non-universal numerical factors in the TED. In
this regularisation all zeros in Eq.~(\ref{EDFexp}) are
understood to be substituted by $a_0$.
Suppressing this argument in the formulas, we obtain the following
expression for the second-order correction to the local (tip)
Green's function (diagram Fig.~\ref{fig3} with an appropriate
arrangement of the spin indices):
\begin{eqnarray} \label{bareint}
 &\,& g^{-+}_{s>}(\omega) = -
\gamma^2 4 U_0^2 \Theta(\omega) g_{0}^{++}(\omega)
 g_{0}^{--}(\omega) \nonumber \\ &\times&
 \int d \epsilon_1 d \epsilon_2 \,
 g_{0}^{-+}(\omega-\epsilon_1) \nonumber
 \\ &\times&  g_{0}^{-+}(\epsilon_1+\epsilon_2)
 g_{0}^{--}(\epsilon_2) g_{0}^{++}(\epsilon_2)
 G_{0}^{+-}(\epsilon_2) \;
\end{eqnarray}
where the local Green's functions in the lead ($G$'s) are understood
to be the same as in the host but with energy measured from $-V$.
To evaluate this diagram, it is useful to remember that
\begin{eqnarray} \nonumber
g_{0}^{++}(\omega) g_{0}^{--}(\omega) = - |g^{R}_{0}(\omega)|^2 =
- 4 \sin^2 \left[(k_F + \omega) a_0\right] \; .
\end{eqnarray}
As we restrict our considerations to energy scales much smaller
than the Fermi energy, we can regard these products essentially
constant. The same is justified for other Green's functions
except, of course, the step functions appearing in front of
the off-diagonal Keldysh functions. Then, calculating energy
integrals in Eq.~(\ref{bareint}), we obtain the following result
for the TED:
\begin{eqnarray} \nonumber
 n_{s>}(\omega)=-ig^{-+}_{s>}(\omega)= C_0 \, \gamma^2 \,  U_0^2 \,
 \Theta(\omega)\Theta(V - \omega) (\omega-V)^2/2 \;
\end{eqnarray}
(with the non-universal constant $C_0 = 4^5 \sin^{10} (k_F a_0)$).

Note that in accordance with the discussion in Section \ref{secII}
no particles can have energy exceeding $V$ at the second order in
the interaction constant. Such particles, however, appear if one
takes into account processes of higher order. The corresponding
diagrams can easily be constructed, see Fig.~\ref{fig3}. So, at
the fourth order in the Hubbard $U$ we found
\begin{eqnarray} \nonumber
 g^{-+}_{s>}(\omega) = i \, C_0^\prime \, \gamma^4 \, U_0^4 \, \Theta(\omega)
 \Theta(2V - \omega) (\omega-2V)^4/24 \;
\end{eqnarray}
($C_0^\prime$ is again a non-universal numerical pre-factor).
It is not difficult to calculate the particle spectrum for arbitrary $\omega$.
In the energy window between $V(n-1)$ and $Vn$ ($n=1,2,...$) it is dominated
by the term of the order $[\gamma U_0]^{2n}$ and shows
$(nV - \omega)^{2n}$ decay.

Another interesting local model is the local phonon model,
where the electron density operator is coupled to a local oscillator
at the tip $x=0$.
In the chiral formulation the Hamiltonian of the problem is
\begin{eqnarray} \label{phononH}
 H &=& i \int \, dx \, \psi^\dag(x) \partial_x
 \psi(x) + \Omega b^\dag b \nonumber \\
 &+& \lambda \psi^\dag(0) \psi(0) (b^\dag + b) \,
\end{eqnarray}
where $b^\dag$ and $b$ are the creation and annihilation
operators of the local phonon. The chirality of fermions takes
care of the reflecting boundary condition at the origin (see
e.~g.~Ref.~\cite{fg}) where the tunnelling processes take place.
In the anti-adiabatic approximation (justified for high-frequency
phonons, $\Omega\gg\lambda/a_0$), this Hamiltonian can be solved
via the Fr\"ohlich transformation \cite{frolich}
\begin{eqnarray} \label{Qtrafo}
 H' = e^{Q} H e^{-Q} \, ,
\end{eqnarray}
with
\begin{eqnarray} \label{Qdef}
 Q = \frac{\lambda}{\Omega} \psi^\dag(0) \psi(0) (b^\dag - b) \, .
\end{eqnarray}

Simple transformations of the field operators, which we omit here,
show that braiding relations are satisfied for this particular
model and the braiding phase $\varphi$ defined in
Eq.~(\ref{braiding}) is equal to:
\begin{eqnarray} \label{sinphase}
 \varphi(t) = \pi + \frac{1}{2} \left(\frac{\lambda}{\Omega a_0}\right)^2
 \sin [\Omega t] \, .
\end{eqnarray}
Using this result when evaluating Eqs.~(\ref{com}) and (\ref{Rdef})
we obtain for the TED:
\begin{widetext}
\begin{eqnarray}
n_>(\omega) &=& -2 i \gamma^2 \int\limits_{-\infty}^{\infty} \,
dt \, \frac{e^{i(\omega-V)t}}{(t+i \alpha)^2} \nonumber
\int\limits_0^\infty \, d\tau_1 \, \int\limits_0^\infty \,
d\tau_2 \, e^{i\omega(\tau_1+\tau_2)}(1-e^{i (\lambda/\Omega
a_0)^2 \sin[\Omega \tau_1]/2})
(1-e^{-i  (\lambda/\Omega a_0)^2 \sin[\Omega \tau_2]/2}) \nonumber \\
&\times& \frac{\tau_1 \tau_2}{(\tau_1+\tau_2+t+i\alpha)(\tau_1+t+i
\alpha)(\tau_2+t+i \alpha)} \, . \nonumber
\end{eqnarray}
\end{widetext}
Expanding in powers of $\lambda/\Omega a_0$ the evaluation of
these integrals yields the main contribution to the TED
of the form
\begin{eqnarray} \label{phononTED}
 n_>(\omega) \approx \frac{\pi \gamma^2}{2} \left( \frac{\lambda}{\Omega a_0}
 \right)^4 \Theta(V-\omega) \frac{\omega (V-\omega)}{\Omega^2-\omega^2} \, .
\end{eqnarray}
As expected, the emerging spectrum has a resonant character
having a sharp pole at the oscillator frequency $\Omega$. Under
appropriate conditions, the Fr\"ohlich transformation can also be
used for solving the bulk electron-phonon interaction where the
braiding relations still persist. We shall not pursue this issue
further in this paper.

\section{Luttinger liquid model} \label{luttinger}
Staying with 1D hosts, we now move on to the next level of difficulty
and discuss the bulk interactions.
The simplest model here is the spinless Luttinger liquid model \cite{haldane}.
Subject to minor modifications (see below) the LL results will also be
applicable to quantum wires and SWNTs.
The interacting term in the host Hamiltonian then is
\begin{equation} \nonumber
H_i = \int \, dx \, dy \, \psi^\dag(x)
\psi^\dag(y) U(x-y) \psi(y) \psi(x) \, .
\end{equation}
where $U(x)$ is the interaction potential.

Although this model can be exactly solved,
it is instructive to start with the perturbative expansion.
Diagram a) in Fig.~\ref{fig3} still represents the only
non-vanishing contribution to the TED above the Fermi edge.
Compared to Eq.~(\ref{bareint}), the analytic expression for this diagram is
slightly more complicated since all participating Green's
functions acquire spatial dependence.
Assuming local interaction, $U(x-y)\to U_0\delta(x-y)$, we write
\begin{widetext}
\begin{eqnarray} \label{bareint1}
 g^{-+}_>(\omega) &=& - \gamma^2 U_0^2 \Theta(\omega) \int_{a_0}^\infty \,
 dx_1 \, dx_2 \, g_0^{++}(x_2,a_0;\omega)
g_0^{--}(a_0,x_1;\omega) \int d \epsilon_1 d
\epsilon_2 \, g_0^{-+}(x_1,x_2;\omega-\epsilon_1) \nonumber \\
 &\times& g_0^{-+}(x_1,x_2;\epsilon_1+\epsilon_2)
 g_0^{--}(x_1,a_0;\epsilon_2)
 g_0^{++}(a_0,x_2;\epsilon_2) G_0^{+-}(a_0,a_0;\epsilon_2) \; .
\end{eqnarray}
We use the same regularisation as in Section \ref{seclocalinteraction}
(that is why the lower boundary of the $x$-integration is the lattice
constant $a_0$).
Substituting the expressions for the corresponding bare Green's functions
into this formula one obtains the following result:
\begin{eqnarray}                 \label{LLfactorize}
g^{-+}_>(\omega) &=& i \gamma^2 U_0^2 \, 8 C_0 \, \Theta(\omega)
\Theta(V-\omega) \int_\omega^{V} \, d\epsilon_1
\int_{-V}^{-\epsilon_1} \, d \epsilon_2 \, |
H(\epsilon_1,\epsilon_2,\omega)|^2 \, ,
\end{eqnarray}
where $C_0$ is a non-universal numerical constant and the
function $H$ is defined as
\begin{eqnarray} \nonumber
H(\epsilon_1,\epsilon_2,\omega) = \frac{1}{2} \int_0^\infty \, dx
\, e^{i(\omega-\epsilon_2)x} \cos[(2 \epsilon_1 + \epsilon_2 -
\omega)x]  = -\frac{1}{4i}
\frac{\epsilon_2-\omega}{\epsilon_1(\epsilon_1 + \epsilon_2 -
\omega)} \, .
\end{eqnarray}
\end{widetext}
We are ignoring terms containing rapidly oscillating integrands
which are negligible on energy scales much smaller than $E_F$.

Performing the last energy integrals we obtain
for the TED above the Fermi edge
\begin{eqnarray} \nonumber
n_>(\omega) = \gamma^2 U_0^2 \, \Theta(\omega) \Theta(V - \omega) C_0 \Big\{
\frac{V-\omega}{\omega} + F_0\Big(\frac{\omega}{V}\Big)
\Big\} \, ,
\end{eqnarray}
where $F_0(x)$ is a function regular at $x=0$ and vanishing for
$x>1$. While the TED vanishes smoothly in the vicinity of $\omega=V$,
there is a sharp singularity towards the Fermi edge.

Since the Luttinger liquid model is solvable for arbitrary
interaction strength, \cite{haldane}, it is possible to go beyond
the perturbative expansion. In particular, the four-point
correlation function in Eq.~(\ref{com}) can be calculated exactly.
We remind the reader that the system under consideration is a
half-infinite spinless LL with an open boundary across which
tunnelling processes occur. Using the standard bosonization
scheme for open boundary LLs (technical details can be found in
the literature, see \cite{fg,book}), we write
\begin{equation} \label{bosrepr}
 \psi(x=0,t) = (2 \pi a_0)^{-1/2} \exp [i \phi(x=0,t)/\sqrt{g}] \,
\end{equation}
where $g$ is the LL parameter, defined by $g=(1+4 U_0/\pi)^{-1/2}$.
The Gaussian chiral Bose field $\phi(x,t)$ has been rescaled
and is governed by the LL Hamiltonian
\begin{eqnarray}
H_{LL}[\phi] = \frac{1}{4 \pi} \int_{-L}^{L} \, dx \, (\partial_x \phi)^2 \, .
\end{eqnarray}
The Bose field is periodic in $2L$, where $L$ is the system size.
In order to obtain the correct analytic properties of correlation
functions we first work with a system of a finite length $L$,
which makes the energy quantisation equal to $\epsilon_0 =
\pi/L$, and then send $L$ to infinity. Using representation
(\ref{bosrepr}) we obtain the four-point correlation function in
question:
\begin{eqnarray} \label{GF}
{\cal K}^{-+}_{+-}(t_1,t_2; t_3,t_4)= (2 \pi b)^{-2}
\mbox{sgn}(t_2-t_3) \mbox{sgn}(t_1-t_4) \nonumber \\
\times \Big[\frac{F(|t_2-t_3|)F(|t_1-t_4|)F^2(0)}{F(t_2-t_1)F(t_2-t_4)F(t_3-t_1)F(t_3-t_4)}
\Big]^{1/g} \, ,
\end{eqnarray}
where the function $F(t)$ is defined by
\begin{eqnarray} \nonumber
 F(t)=1-e^{i\epsilon_0 (t+i \delta)} \, .
\end{eqnarray}
In the thermodynamic limit  $L\rightarrow \infty$, we expand Eq.~(\ref{GF})
in powers of $\epsilon_0$.
Then Eq.~(\ref{com}) can be brought into the following form
\begin{widetext}
\begin{eqnarray} \label{genLLres}
n(\omega) &=& -2 i \gamma^2 e^{-i \pi/g} \cos^2(\pi/2g) \int dt
\, \frac{e^{i(\omega-V)}}{(-t-i \alpha)^{1+1/g}} \int_0^\infty
\int_0^\infty \, d\tau_1 d\tau_2 e^{i\omega(\tau_1+\tau_2)}
\nonumber \\ &\times& \frac{(\tau_1
\tau_2)^{1/g}}{[(-\tau_1-\tau_2-t-i\alpha) (-\tau_1 -t -i \alpha)
(-\tau_2-t -i\alpha)]^{1/g}} \, , \nonumber
\end{eqnarray}
\end{widetext}
where we used the fact that the LL field operators satisfy the
braiding relations, (\ref{braiding}), with the braiding phase
function
\begin{eqnarray}
\varphi(t) = \frac{\pi}{g}\, \mbox{sign} (t) \, .
\end{eqnarray}
Now we can perform the $\tau$-integrations
using the integral representation
\begin{eqnarray} \nonumber
(\tau_1+\tau_2-t-i\alpha)^{-1/g} &=& \frac{e^{i \pi/2g}}{\Gamma(1/g)}
\int_0^\infty dp  \, p^{1/g-1} \\
&\times& e^{-i(\tau_1+\tau_2-t)p} \, . \nonumber
\end{eqnarray}
Then
\begin{eqnarray}
 &~& n_>(\omega) = 2 i \gamma^2 e^{-i \pi/g} \cos^2(\pi/2g) \\
 &\times&  \int\limits_{-\infty}^\infty dt \,
 \frac{e^{i(\omega-V)}}{(-t-i \alpha)^{1+1/g}}
\nonumber \\ &\times& \Gamma^2(1/g+1)
\int_0^\infty \, dp \, p^{1/g-3} e^{i p t} \Psi^2(1/g,0; -i (p+\omega)t) \, , \nonumber
\end{eqnarray}
where $\Gamma$ and $\Psi$ are the gamma function and the Tricomi confluent
hyper-geometric function, respectively \cite{bateman}.
Deforming the $t$-integration contour from the real axis to
the contour around the branch cut of the function $\Psi$ we arrive at a
more convenient representation for the TED:
\begin{eqnarray} \label{finalint}
n_>(\omega) = A(g) \int_0^{V-\omega} dE \, \frac{E^{1/g-1}}{(E+\omega)^2}
F_V(E+\omega) \, ,
\end{eqnarray}
where the spectral function is given by
\begin{eqnarray} \label{psiintegral}
F_V(p) = 2 \, \mbox{Im}\, e^{-i\pi/g} \int_\alpha^\infty d\xi \, \xi^{-1/g-1}
e^{-(V-p)\xi} \nonumber \\
\times \Psi^2(1/g,0,-p \xi+i\alpha) \, ,
\end{eqnarray}
and the pre-factor is
\begin{eqnarray} \nonumber
A(g) = \Gamma^2(1/g+1) \gamma^2 \alpha^{2/g} \frac{\cos^2[\pi/ 2 g]}{2 \pi
a_0^2 \Gamma(1/g)} \; .
\end{eqnarray}
In the limit of small energies, $\omega/V \ll 1$, the function
$\Psi$ is nearly constant and the limiting form
of the spectral function can easily be established:
\[
F_V(p) \approx \frac{2 \pi}{\Gamma^3(1+1/g)} (V-p)^{1/g} \, .
\]
Therefore we obtain the following asymptotic form
for the TED in vicinity of the Fermi edge
\begin{eqnarray} \label{cresult}
n(\omega) \approx C_2 (\lambda+1)^2 (\omega/V)^{\lambda} \, ,
\end{eqnarray}
where $\lambda = 1/g-2$ and $C_2$ is a
numerical pre-factor regular at $\lambda=-1$.

We investigated the behaviour of the TED near the upper threshold
$\omega=V$ by numerical evaluation of Eq.~(\ref{finalint}). The
result is given by another power law:
\begin{equation} \label{exprelation}
 n(\omega) \sim (V-\omega)^\nu \,\, , \,\, \nu=\lambda+2 \, .
\end{equation}

In the limit of weak interactions, $g\rightarrow 1$, the exponent
$\lambda$ approaches the perturbative result. Eq.~(\ref{cresult})
can be regarded as consisting of two factors. The first factor is
the universal $1/\omega$ divergence inherent to all interacting
1D systems. The second factor reflects power-law renormalisations
occurring in the LLs and, in particular, contains the local
density of states (LDOS) of the primary electron. The latter
object is suppressed for repulsive interactions ($g<1$). So also
the TED singularity is suppressed. At $g_c=1/2$ the LDOS
suppression effectively wins over and the singularity disappears.
It is noteworthy that in the case when the interaction constant is
smaller than this critical value, the TED has a maximum between
the upper and lower thresholds $\omega=0$ and $\omega=V$, as it
is vanishing towards both limits.

In real systems the tunnelling amplitude $\gamma$ is small. The
secondary current is already calculated in the next-to-leading
(compared to to the primary current) approximation in $\gamma$.
Still, it is a valid question to ask what happens at higher
orders in the tunneling amplitude. For the case when the TED is
divergent at the Fermi level ($\lambda<0$), a small energy scale
$\omega^*$ may emerge below which these higher order
contributions become important. We do not currently have full
answer to this question. Our preliminary calculations indicate,
however, that the most divergent 4-th order diagram is the one
for which Fig.~\ref{fig4} acts as self-energy. This would lead to
the estimate $\omega^*\sim V\gamma^{-2/\lambda}$. This issue
deserves further investigation.

Thus far we have discussed the spinless LL. The above results can
be straightforwardly generalised to spinful systems such as
quantum wires. Due to the spin-charge separation the LDOS at the
end of the wire is then given by the same expression as in the
spinless case with $1/g$ substituted by $(1/g_c + 1/g_s)/2$,
where $g_{c,s}$ are the interaction constants in the charge and
spin sectors, respectively \cite{book}. Since the tunnelling
processes conserve spin, the LDOS exponent is simply carried over
to the TED, so that $\lambda$ in Eq.~(\ref{cresult}) should be
changed to:
\begin{equation} \label{lwires}
\lambda\rightarrow\lambda=(1/g_c+1/g_s)/2-2\,.
\end{equation}
The relation between the lower and upper threshold exponents,
Eq.~(\ref{exprelation}), is still valid. When the spin SU(2)
symmetry is preserved, it forces $g_s=1$. The residual
spin-backscattering interaction renormalises to zero under
renormalisation group transformations (see \cite{book} for a
recent review on 1D physics and further references). However, the
presence of an irrelevant operator should probably cause
multiplicative logarithmic corrections to our power-law formula
(\ref{cresult}). We have not attempted to calculate these
correction though we expect that this can be done by standard
methods \cite{affleck}.

\section{Tunnelling from 2D systems: pure limit} \label{pure}
In view of applications to MWNTs, we now consider tunnelling from
a 2D electron system in tip geometry. Theoretically, this is a
more complicated situation than in 1D. In some cases, relevant
non-perturbative techniques exist, in other cases we shall present
perturbative results. In this Section we discuss interacting
2D electron systems free from impurities (pure limit).

From the mathematical point of view we still have to calculate the same
diagram [(a) in Fig.~\ref{fig3}].
To accomplish this task in 2D it is convenient to change over to
the momentum representation.
The host Hamiltonian is $H[\psi]=H_0[\psi]+H_i[\psi]$,
where
\begin{equation}\label{H02D}
H_0[\psi]=\int\frac{d\vec{p}}{2\pi}\xi(\vec{p})\psi_{\vec{p}}^{\dag}
\psi_{\vec{p}}
\end{equation}
describes free 2D electron gas with dispersion relation $\xi(\vec{p})$
and $\psi_{\vec{p}}$ is the Fourier transform of the electron field operator.
The interaction term has the standard form. In real space:
\begin{equation}\label{Hi2D}
H_i = \int \, d\vec{r} \, d\vec{r'} \, \psi^\dag(\vec{r})
\psi^\dag(\vec{r'}) U_0(|\vec{r}-\vec{r'}|) \psi(\vec{r'}) \psi(\vec{r}) \, ,
\end{equation}
where $U_0(r)$ is the interaction potential to be specified later.
The relevant diagram is given by the following expression:
\begin{eqnarray}                    \label{wholecontrib}
g^{-+}_>(\omega) &=& \int \, \frac{d^2 \vec{q_1}}{(2\pi)^2} \, \int
\frac{d^2\vec{q_2}}{(2\pi)^2} \, \int \frac{d \Omega}{2 \pi}
\int \frac{d^2\vec{p}}{(2\pi)^2} \, \nonumber
\\ &\times&  g^{--}_0(\omega,\vec{p}-\vec{q_1})
g^{-+}_0(\omega-\Omega,\vec{p})
\\ \nonumber &\times&
U_0(q_1)\Pi(\Omega;\vec{q_1},\vec{q_2})U_0(q_2)
g^{++}_0(\omega,\vec{p}-\vec{q_2})
\end{eqnarray}
Here $U_0(q)$ is the Fourier transform of the interaction potential,
$\Pi(\Omega;\vec{q_1},\vec{q_2})$ stands for the inner polarisation
bubble as in Fig.~\ref{fig3} (b) with vertex indices as in (a):
\begin{eqnarray}                    \label{bubbledef}
 \Pi(\Omega;\vec{q_1},\vec{q_2}) &=& \int \frac{d \epsilon}{2 \pi}
\int \frac{d^2
 \vec{p}}{(2\pi)^2} \,   g^{--}_0(\epsilon,\vec{p}+\vec{q_1}) \nonumber
  \\
 &\times& G^{-+}_0(\epsilon)  g^{++}_0(\epsilon,\vec{p}+\vec{q_2})
 g^{-+}_0(\epsilon+\Omega,\vec{p}) \, . \nonumber
\end{eqnarray}
Without the tunnelling insertions, this would be a Keldysh analogy
to the standard 2D Lindhard functions \cite{benard}.
We define $\vec{p} = P_\omega \vec{n}_\phi$ with a unit 2D vector
$\vec{n}_\phi$ in direction parametrised by the angle $\phi$.
The quantity $P_\omega$ gives the on-shell magnitude of the
momentum $\vec{p}$ and, for a quadratic dispersion relation,
satisfies the following equation,
\begin{eqnarray}                      \label{Pdef}
 \frac{P_\omega^2}{2 m} = \frac{p_F^2}{2 m} + \omega \, ,
\end{eqnarray}
where $p_F$ is the Fermi momentum and $m$ is the effective mass.

We change over from the momentum integration to the energy-angle
integration in the standard manner:
\begin{eqnarray} \nonumber
 \int \frac{d^2 \vec{p}}{(2 \pi)^2} \rightarrow \frac{1}{2 \pi} \int d \xi
 \rho(\xi) \int_0^{2 \pi} d \phi \, ,
\end{eqnarray}
where $\rho(\xi)$ is the density of states in the host.
The polarisation bubble can then be written in the form
\begin{widetext}
\begin{eqnarray}
 \Pi(\Omega;\vec{q_1},\vec{q_2}) = \rho_c  \Theta(\Omega-V)
 \int_{-V}^{-\Omega} \, d\epsilon \, \rho(\epsilon + \Omega)
\nonumber \int_0^{2 \pi}
 d\phi g^{--}_0(\epsilon,P_{\epsilon+\Omega} \vec{n}_\phi + \vec{q_1})
g^{++}_0(\epsilon,P_{\epsilon+\Omega} \vec{n}_\phi + \vec{q_2}) \, ,
\nonumber
\end{eqnarray}
\end{widetext}
where $\rho_c$ stands for the constant density of states in the lead.
The diagonal bare Keldysh Green's functions are
\begin{eqnarray} \nonumber
& & g^{--(++)}_0(\epsilon,P_{\epsilon+\Omega}\, \vec{n}_\phi + \vec{q})
\\ \nonumber
&=& \mp
 \left[\Omega + \frac{1}{m}P_{\epsilon+\Omega}\,\vec{n}_\phi \vec{q} +
 \frac{1}{2 m} q^2 \pm i0 \right]^{-1} \; .
\end{eqnarray}
In this expression we have taken into account the fact that $\epsilon < 0$
if we are only interested in $\omega>0$ part of $g^{-+}$.
The off-diagonal Keldysh function is
\begin{eqnarray} \nonumber
 g^{-+}_0(\omega-\Omega,\vec{p}) = i 2 \pi n_F(\vec{p})\, \delta(
 \omega-\Omega-\xi_{\vec{p}} ) \, ,
\end{eqnarray}
where $n_F(\vec{p})$ is the Fermi momentum distribution
function. Using these expressions and relation (\ref{tokexp1})
between the Green's function and the TED we obtain a formula
similar to Eq.~(\ref{LLfactorize}) for the 1D case, namely
\begin{eqnarray}                          \label{nbol}
 n_{>}(\omega) = \frac{\rho_c}{2 \pi} \int_\omega^V \, d \Omega \,
 \rho(\omega-\Omega) \int_{-V}^{-\Omega} \, d\epsilon \, \rho(\epsilon+\Omega)
 \\ \nonumber \times
 \int_0^{2 \pi} \, d\phi \, d\phi' |F_{\epsilon}(\Omega,\phi-\phi')|^2 \,
 ,
\end{eqnarray}
where the function $F$ is defined by
\begin{eqnarray}                                 \label{woher}
 & & F_{\epsilon}(\Omega,\phi) = \frac{1}{4 \pi^2} \int_0^\infty
 dq \, q \, U_0(q) \int_0^{2 \pi} d\theta
 \nonumber \\ \nonumber
 &\times& [\Omega + \frac{1}{m} P_{\epsilon+\Omega} q
 \cos(\theta-\phi/2) + \frac{1}{2m} q^2 + i0]^{-1}
 \\ \nonumber
 &\times&
 [\Omega + \frac{1}{m} P_{\epsilon+\Omega} q
 \cos(\theta+\phi/2) - \frac{1}{2m} q^2 + i0 ]^{-1} \, .
\end{eqnarray}

To proceed, we now need to specify the interaction potential. It
is natural to start with a contact interaction term,
$U(\vec{r}-\vec{r'}) = U_0\,\delta(\vec{r}-\vec{r'})$ or
$U_0(q)=U_0$ in the momentum space. However, an unexpected
technical difficulty arises. If we follow the standard practice
and linearise the electron spectrum in the vicinity of the Fermi
surface, then the momentum integration in Eq.~(\ref{woher}) can
be easily done but the remaining angle integration diverges at
$\varphi=\pm\pi$. This latter singularity comes from large
momenta in the previous momentum integral and corresponds to the
special case of hot hole back-scattering. To obtain a finite
result for the TED one either needs to take into account the
decay of $U_0(q)$ for large $q$ or the non-linearity of the
electron spectrum. Accordingly, we present two different
calculations in the rest of this Section.

{\bf (i)} Contact potential: $U_0(q)=U_0$.

First we scale the integration variable $q
\rightarrow \Omega q$ and introduce dimensionless quantities
$\gamma_{1,2}= P_{\epsilon
\pm \Omega}/ p_F$ and $\kappa = \Omega/4 E_F$.
Then,
\begin{eqnarray}                     \label{transformed}
 & &  F_{\epsilon}(\Omega,\phi) = \frac{U_0}{4 \pi^2 v_F^2}
 \int_0^\infty dq q \int_0^{2 \pi} d\theta
 \nonumber \\ \nonumber
 &\times& [1+ \gamma_1 q \cos(\theta - \phi/2) + \kappa q^2 + i0]^{-1}
 \\ \nonumber
 &\times& [1+ \gamma_2 q \cos(\theta + \phi/2) - \kappa q^2 + i0]^{-1} \, .
\end{eqnarray}
Notice that $\gamma$'s approach unity whereas $\kappa$ linearly
tends to zero in the limit of small $\Omega$.
Furthermore, if we set $\kappa=0$, the angle integration is still divergent.
Therefore we set $\gamma_{1,2}=1$ and keep $\kappa$ finite.
As a next step we expand the integrand in Eq.~(\ref{transformed})
using the standard formula \cite{gelfand},
\begin{eqnarray}               \label{gshilov}
  \frac{1}{f(x)+i 0} = {\cal P}
 \frac{1}{f(x)} - i \pi \delta(f(x)) \, ,
\end{eqnarray}
where ${\cal P}$ denotes the principal value. This expansion
produces three contributions: the first one contains the product
of two principal parts and remains regular for $\Omega
\rightarrow 0$, the second one contains products of one principal
part and one delta function and is identically zero in the low
energy regime, and the third contribution, $F^\delta_{\epsilon}$,
containing a product of two delta functions is responsible for
the divergency or the angle integral. Setting $v_F=1$ for the
rest of this Section, we write
\begin{widetext}
\begin{eqnarray}
 F^\delta_{\epsilon}(\Omega,\phi) = \frac{U_0}{4 \pi^2}
 \int_0^\infty dq q \int_0^{2 \pi} d\theta
 \nonumber
 \delta[ 1+ q \cos(\theta - \phi/2) + \kappa q^2 ]
\delta[ 1+ q \cos(\theta + \phi/2) - \kappa q^2 ] \, .
\end{eqnarray}
The evaluation of the angle integration leaves us with
\begin{eqnarray}
 F^\delta_{\epsilon}(\Omega,\phi) = \frac{U_0}{4 \pi^2}
 \int_0^\infty dq \, q \,[q^2-(1+\kappa q^2)^2]^{-1/2}
\nonumber \nonumber
 \delta[ 1-(1+\kappa q^2)\cos \phi + \sin \phi
\sqrt{q^2-(1+\kappa q^2)^2}-\kappa q^2] \, .
\end{eqnarray}
\end{widetext}
The argument of the delta function is zero for
\begin{eqnarray} \nonumber
 q_0^2 = \frac{1}{2 \kappa^2} \sin^2(\phi/2)
\Big( 1-\sqrt{1-16 \frac{\kappa^2}{\sin^2 \phi} } \Big) \, .
\end{eqnarray}
Therefore the remaining momentum integration yields
\begin{eqnarray}                     \label{calculdF}
 F^\delta_{\epsilon}(\Omega,\phi) &=&
\frac{U_0}{2 \pi^2} \Big[ \sin \phi + \cos \phi \Big( \sin \phi
\nonumber \\ \nonumber &-& \sqrt{\sin^2 \phi - 16 \kappa^2} \Big)
- 4 \kappa \sin \phi \Big]^{-1} \, .
\end{eqnarray}
Plugging this expression into Eq.~(\ref{nbol}) and taking into account
that for small $\Omega$ (and hence small $\kappa$) the main contribution
to the angle integration originates for $\phi$ close to $\pm\pi$
one finds for the following limiting form for the TED:
\begin{eqnarray} \nonumber
 n_>(\omega) \approx \frac{U_0^2}{8 \pi^4} \rho_c \nu^2 E_F V
 \ln \left(\frac{V}{\omega}\right) \, ,
\end{eqnarray}
where $\nu = \rho(0)$.

{\bf (ii)} Yukawa potential: $U_0(q) = 4 \pi e^2/\sqrt{q^2+\lambda^2}$.

Here $1/\lambda$ is the screening length. As $U_0(q)$ now tends
to zero for large $q$ (which is, of course, always the case in
real systems), we can linearise the dispersion relation. Then the
function $F(\Omega,\phi)$ becomes independent of the energy
variable $\epsilon$ and simplifies considerably:
\begin{eqnarray}                       \label{newF}
 F(\Omega,\phi) = \frac{1}{4 \pi^2} \int_0^\infty \, dq \, q \,
U_0(\Omega q) \, I(q,\phi) \, ,
\end{eqnarray}
where $q$ has again been scaled by $\Omega$ and the function $I(q,\phi)$
is defined by
\begin{eqnarray}                        \label{Idef}
 I(q,\phi) &=& \int_0^{2 \pi} \, d\theta \,\Big[(1+ q \cos(\theta-\phi/2)+i0)
 \nonumber \\ &\times& (1+ q \cos(\theta+\phi/2)+i0)\Big]^{-1} \, .
\end{eqnarray}
This integral can be calculated and is real
\begin{eqnarray}                      \label{I1}
 I(q,\phi) = \frac{2 \pi}{[1-q^2 \cos^2(\phi/2)]\sqrt{1-q^2}} \, ,
\end{eqnarray}
for $q<1$ but contains an imaginary part
\begin{eqnarray}                      \label{I2}
 I(q,\phi) &=& -i \frac{2 \pi}{[1-q^2 \cos^2(\phi/2)]\sqrt{q^2-1}}
  \\ \nonumber
  &-& \frac{\pi^2}{\sqrt{q^2-1} \cos(\phi/2)} \delta[q-1/\cos(\phi/2)] \, .
\end{eqnarray}
for $q>1$, as incoherent particle production takes place in the
latter regime. Using Eqs.~(\ref{I1}) and (\ref{I2}) we compute the
momentum integral in Eq.~(\ref{newF}) which yields
\begin{eqnarray}                         \label{finalF}
 F(\Omega,\phi) &=& - \frac{2 e^2}{|\sin(\phi/2)|
\sqrt{\Omega^2+\lambda^2 \cos^2(\phi/2)}} \nonumber \\
&\times& \mbox{arccot} \Big[ \frac{\sqrt{\Omega^2 + \lambda^2
\cos^2(\phi/2)}}{\lambda |\sin(\phi/2)|} \Big] \, .
\end{eqnarray}
Unfortunately we were not able to perform the last remaining angle
integration in Eq.~(\ref{nbol}) in a closed form. However, to
analyse the TED close to the Fermi surface $\omega\rightarrow 0$
one merely requires the knowledge of $|F(\Omega,\phi)|^2$ for
small $\Omega$'s. The latter can be easily read off
Eq.~(\ref{finalF}) for $\Omega \ll \lambda$. There are two
different regions in the parameter space $(\phi, \Omega)$: for
most angles,  $|\phi\pm \pi|> 2\Omega/\lambda$, we find
\begin{eqnarray} \nonumber
|F(\Omega,\phi)|^2 \approx \frac{4 e^4}{\lambda^2} \frac{\phi^2}{\sin^2 \phi}
\end{eqnarray}
whereas for large scattering angles  $|\phi\pm \pi|< 2\Omega/\lambda$
one obtains a singular in $\Omega$ behaviour,
\begin{eqnarray} \nonumber
 |F(\Omega,\phi)|^2 \approx \frac{\pi^2 e^2}{\Omega^2} \, .
\end{eqnarray}
The latter region, physically corresponding to hot hole
back-scattering processes, is responsible for the leading
contribution to the remaining integrals. With logarithmic
accuracy we thus obtain the following threshold behaviour of the
TED,
\begin{eqnarray}                     \label{logresult}
 n_>(\omega) \approx \rho_c \nu^2 \frac{2 \pi e^4}{\lambda}
\ln \left[ \frac{\mbox{max}(V,\lambda)}{\omega} \right] \, .
\end{eqnarray}
Clearly the same logarithmic singularity will persist for other
functional forms of $U_0(q)$ as long as it decays for large $q$.
(We note that in the purely Coulomb case, $\lambda=0$, an
additional divergence at small $q$ is introduced. Calculation
shows that all the divergences cancel exactly in this special
case with the second order result for the TED being identically
zero for all $\omega>0$. The physical meaning of this observation
escapes us.)

The net result is that we find a logarithmic singularity in the TED
for 2D electron systems at the second order in the interaction.
Such behaviour is in apparent contradiction with Ref.~\cite{gadzuk}
where a stronger singularity was found.
These differences may be due to the fact that
Gadzuk and Plummer analysed a different set-up with a three-dimensional host
and a 2D emitting surface (as opposed to our set-up with a 2D host and tip
emission), or their stronger singularity may be an artifact of the
low-density approximation.
This issue as well as the role of higher-order scattering
processes deserve
more intensive study and will be discussed elsewhere.
In view of applications to MWNTs, however, there are two
more immediate questions: the role of the dynamical Coulomb screening
and of the disorder potential.
Both require a non-perturbative approach and will be discussed
in the next Section.

\section{Tunnelling from disordered 2D systems} \label{disorder}
Contrary to the SWNTs, MWNTs can be described by
the LL theory only in some special cases when the
radii of outer shells are not too large \cite{eggeronly}.
Otherwise their physics is best described in
terms of a disordered 2D electron liquid with Coulomb interaction.
Such systems have been extensively studied in the 80's
using non-linear sigma model and renormalisation group
\cite{finkel,castellani,belitz}.
Recently they attracted more attention due to the discovery
of a possible metal-insulator transition in MOSFETs
\cite{kravchenko}, which stimulated further theoretical research.
Kamenev and Andreev (KA)  \cite{kamenev} have recently
examined the Keldysh non-linear sigma model \cite{horbach}
for the Coulomb case (long-range interaction).
In this Section we apply their method to obtain the threshold
behaviour of the TED for tunnelling from
a 2D interacting disordered metal.

The Hamiltonian of the previous Section, composed from (\ref{H02D})
and (\ref{Hi2D}) with $U_0(q)=4\pi e^2/q$, should now be supplemented
by the disorder term
\begin{equation} \label{Hd2D}
H_{d}[\psi]=\int d\vec{r} V(\vec{r})\psi^{\dag}(\vec{r})\psi(\vec{r})
\end{equation}
with a $\delta$-correlated Gaussian disorder potential
$V(\vec{r})$.

The formal procedure of deriving the Keldysh sigma model is
described in the relevant literature (\cite{horbach,kamenev}, see
also \cite{ludwig}) in detail. Therefore we only give a brief
outline here. One starts with the Keldysh $S$-matrix. (This
object is identically unity but it becomes a generating
functional upon introducing auxiliary fields. If one wants to
start with a diagonal bare Keldysh function, the contour is
different from that in Fig.~\ref{contour}, see explanation in
Refs. \cite{horbach,ludwig}). One can then integrate the disorder
out obtaining a non-local in time four-fermion interaction term.
The action is made quadratic in fermions by applying the
Hubbard-Stratonovich transformation to the latter term as well as
to the Coulomb interaction term. Hence there are two decoupling
fields, $Q$ and $\Phi$ in the notation of Ref.~\cite{kamenev}. The
fermions can now be formally integrated out resulting in the
following expression for the generating functional \cite{kamenev}:
\begin{eqnarray}\label{genfunc}
\langle Z \rangle = \int {\cal D} \Phi e^{i \mbox{Tr}
(\Phi^T U_0^{-1} \sigma_1 \Phi)} \int {\cal D} Q e^{i S[Q,\Phi]} \, .
\end{eqnarray}
with the sigma-model action (after Keldysh rotation)
\begin{eqnarray}
i S[Q,\Phi] &=& - \frac{\pi \nu}{4 \tau} \mbox{Tr} \, Q^2 \nonumber \\
&+& \mbox{Tr} \, \ln [G_0^{-1} + \frac{i}{2 \tau_0} Q +
\phi_\alpha \gamma^\alpha + \zeta ] \, , \nonumber
\end{eqnarray}
Here $\tau_0$ is the elastic mean free time and
$U_0$ stands for the unscreened interaction potential.
The decoupling field $Q$ depends on two time variables and
is a 2$\times$2 matrix in the Keldysh space.
The decoupling field $\Phi$ is a Keldysh doublet $(\phi_1, \phi_2)^T$.
The vertex matrices are defined as follows:
\begin{eqnarray} \nonumber
\gamma^1 = \left(  \begin{array}{rr}
                  1 & 0 \\
                  0 & 1
                  \end{array}
           \right)
           \; , \;
\gamma^2 = \sigma_1 = \left( \begin{array}{rr}
                                0 & 1 \\
                                1 & 0
                             \end{array}
                      \right) \, .
\end{eqnarray}
Doing the functional variation of the generating functional,
(\ref{genfunc}), with respect to the auxiliary field $\zeta$
and setting the latter to zero one obtains
the complete set of single-particle Green's functions:
\begin{eqnarray} \nonumber
\left. \frac{\delta}{\delta \zeta}
\langle Z \rangle \right|_{\zeta=0}
= \left( \begin{array}{cc}
G^R & G^K \\
0   & G^A
\end{array} \right) = {\cal G} \, .
\end{eqnarray}
where the function $G^K = (G^R-G^A)(1-2 n(E))$ is
related to the single-particle
energy distribution function $n(E)$ and
$G^{R(A)}$ are  the retarded and advanced
Green's functions, respectively.

We need to calculate of the four-point correlation function
${\cal K}_{+-}^{-+}(t_2,t_3;t_1,t_4)$.
(A calculation of a different four-point function for a
non-equilibrium noise problem has recently been done within a similar
framework in \cite{grabert}.)
This can be achieved by double variation of the generating
functional with respect to the auxiliary field $\zeta$:
\begin{eqnarray} \label{doubleW}
{\cal K}_{+-}^{-+}(t_2,t_3;t_1,t_4) = \int {\cal D}
\Phi e^{i \mbox{Tr} (\Phi^T V_0^{-1} \sigma_1 \Phi)}
\int {\cal D} Q e^{i S[Q,\Phi]} \nonumber \\
\times \left[ W(t_1,t_4)\otimes W(t_3,t_2) + W(t_1,t_2)\otimes
W(t_3,t_4) \right]_* \,
\end{eqnarray}
where
\begin{eqnarray} \nonumber
W(t_1,t_2) = \left[ G_0^{-1} + \frac{i}{2 \tau_0} Q +
\phi_\alpha \gamma^\alpha \right]^{-1} \, .
\end{eqnarray}
A functional integral of this type with a single $W$ function
yields the Green's function matrix ${\cal G}$ and was discussed
in \cite{kamenev}. A product of two and more $W$ functions
produces a complicated object, which is a tensor product
containing all possible time orderings. The operation of
extracting the particular component corresponding to ${\cal
K}_{+-}^{-+}$ we denote by $[...]_*$. At this stage its exact
definition is unimportant and we postpone it to the
end of the Section.

We now evaluate the functional integral with the product
of two $W$-functions in the saddle point approximation
following KA's approach \cite{kamenev}.
In this case
\begin{eqnarray}  \label{Wsaddlpoint}
 W(t,t') = - i \pi \nu \, e^{i k_\alpha(t) \gamma^\alpha} \,
 \Lambda(t-t') \, e^{i k_\alpha(t') \gamma^\alpha} \, ,
\end{eqnarray}
where $\Lambda(t)$ denotes the mean-field non-interacting
Green's function matrix in equilibrium (corresponding to
non-crossing disorder diagrams \cite{agd})
and $k_\alpha(t)$ is a linear functional of $\Phi$.
The corresponding correlation matrix with respect
to averaging over the $\phi$-fields is given by \cite{kamenev}:
\begin{eqnarray} \label{corrmat}
\langle k_\alpha(q,\omega)k_\beta(-q,-\omega) \rangle_\Phi =
\frac{i}{2} {\cal V}_{\alpha \beta} (q,\omega) \, ,
\end{eqnarray}
\begin{eqnarray}    \label{mainVs}
{\cal V}_{\alpha \beta} (q,\omega) &=&
\left( \begin{array}{cc}
{\cal V}^R(q,\omega) & {\cal V}^R(q, \omega) \\
{\cal V}^A(q,\omega) & 0
\end{array} \right) \nonumber \\
{\cal V}^{R(A)}(q,\omega) &=& -
\frac{1}{(d q^2\mp i\omega)^2}\left[
\frac{1}{V_0} + \frac{\nu d
q^2}{d q^2 \mp i \omega}\right]^{-1} \\
{\cal V}^K &=& n_B(\omega) ({\cal V}^R(q,\omega) -
{\cal V}^A(q,\omega)) \, , \nonumber
\end{eqnarray}
where $n_B(\omega)$ is the Bose distribution function
and $d$ is the electron diffusion constant.
Using the saddle point representation in Eq.~(\ref{Wsaddlpoint})
the product of two $W$-functions can be re-written as
\begin{eqnarray}                              \label{4pis}
&~& \langle \left[ W(t_1,t_4) \otimes \nonumber W(t_3,t_2)
\right]_* \rangle_\Phi \\ \nonumber &=& \frac{1}{4} \sum_{\mu_i}
\left[ (\gamma^{\mu_1} \Lambda(t_1,t_4) \gamma^{\mu_4}) \otimes
(\gamma^{\mu_3}
\Lambda(t_3,t_2) \gamma^{\mu_2}) \right]_* \nonumber \\
&\times& \langle p_{\mu_1}(t_1) (p_{\mu_4}(t_4))^* p_{\mu_3}(t_3)
(p_{\mu_4}(t_2))^* \rangle_\Phi \, .
\end{eqnarray}
The operation of taking the correct time ordering is now
separated from of the averaging over the field $\Phi$, which
only affects the objects
\begin{eqnarray} \nonumber
&~& p_\mu(t) = e^{i(k_1(t)+k_2(t))} + \mu
e^{i(k_1(t)-k_2(t))} \\ \nonumber &=& s(t) + \mu \bar{s}(t) =
e^{i \alpha (k_1(t)+ \beta k_2(t))} + \mu e^{i \alpha (k_1(t)-
\beta k_2(t))}  \, .
\end{eqnarray}
The product of four $p_\mu$-operators is a sum of
16 products of different $s(t_i)$-operators.
The latter objects can be evaluated using
the correlation matrix, Eq.~(\ref{corrmat}),
\begin{eqnarray} \label{manyVs}
&& \langle e^{i \alpha_1 (k_1(t_1)+ \beta_1 k_2(t_1))}
e^{-i \alpha_4 (k_1(t_4)+ \beta_4 k_2(t_4))} \nonumber \\
&\times& e^{i \alpha_3 (k_1(t_3)+ \beta_3 k_2(t_3))}
e^{-i \alpha_2 (k_1(t_2)+ \beta_2 k_2(t_2))} \rangle_\Phi \nonumber \\
&=& \exp \Big\{ -\frac{i}{2} \Big[ \alpha_1 \alpha_3 {\cal V}_{13}
(\beta_1,\beta_3)-\alpha_1 \alpha_4 {\cal V}_{14}
(\beta_1,\beta_4) \nonumber \\
&-& \alpha_4 \alpha_3 {\cal V}_{43}(\beta_4,\beta_3)
+ \alpha_4 \alpha_2 {\cal V}_{42}(\beta_4,\beta_2) \\
&-& \alpha_3 \alpha_2 {\cal V}_{32}(\beta_3,\beta_2) -
\alpha_1 \alpha_2 {\cal V}_{12}(\beta_1,\beta_2) +
2 {\cal V}_0(\beta_1,\beta_1) \Big] \Big\} \nonumber \, ,
\end{eqnarray}
where we use the following definition
\begin{eqnarray} \nonumber
{\cal V}_{ij}(\beta_i,\beta_j) = {\cal V}^K(t_i-t_j)
+ \beta_i {\cal V}^A(t_i-t_j) \nonumber \\
+ \beta_j {\cal V}^R(t_i-t_j) \, .
\end{eqnarray}
Taken separately these functions are divergent. Nevertheless
Eq.~(\ref{manyVs}), being rewritten as a function of differences
${\cal V}_{ij}(\beta_i,\beta_j)-{\cal V}_{ii}(\beta_i,\beta_i)$,
can be made convergent. In this case a special selection rule has
to be fulfilled,
\begin{eqnarray} \label{selrule}
&& \alpha_1 \alpha_3 {\cal V}_0(\beta_1,\beta_3) -
\alpha_1 \alpha_4 {\cal V}_0(\beta_1,\beta_4) \nonumber \\
&-& \alpha_4 \alpha_3 {\cal V}_0(\beta_4,\beta_3) +
\alpha_4 \alpha_2 {\cal V}_0(\beta_4,\beta_2) \nonumber \\
&-& \alpha_3 \alpha_2 {\cal V}_0(\beta_3,\beta_2) -
\alpha_1 \alpha_2 {\cal V}_0(\beta_1,\beta_2) \nonumber \\
&+& \frac{1}{2} \sum_{j=1,..,4} {\cal V}_0(\beta_j,\beta_j) = 0
\, ,
\end{eqnarray}
otherwise (\ref{manyVs}) is divergent. Using Eqs.~(\ref{mainVs})
one can calculate the asymptotic behaviour of ${\cal
V}_{ij}(\beta_i,\beta_j)-{\cal V}_{ii}(\beta_i,\beta_i)$ for
large time differences (to calculate the limiting Fermi edge
asymptotics of the TED this knowledge is sufficient as we shall
see shortly). Fortunately, the limiting form turns out to be
independent of the arrangement of the $\beta$-indices, with the
result
\begin{eqnarray} \nonumber
{ \cal V}_{ij}(\beta_i,\beta_j)-{\cal V}_{ii}(\beta_i,\beta_i)
\approx i \frac{e^2}{4 \pi f} \ln^2 \left| \frac{f^2 (t_i-t_j)}{d} \right| \, ,
\end{eqnarray}
where $f=2 \pi e^2 \nu d$.
The last average in Eq.~(\ref{4pis}) is then given by
\begin{eqnarray}
\langle p_{\mu_1}(t_1) (p_{\mu_4}(t_4))^*
p_{\mu_3}(t_3) (p_{\mu_4}(t_2))^* \rangle_\Phi \nonumber \\
= \left(\prod_{j}(1+\mu_j)\right)\langle
s(t_1)s^*(t_4)s(t_3)s^*(t_2)\rangle_\Phi \, . \nonumber
\end{eqnarray}
Obviously, the above is nonzero only if for all $i$ $\mu_i=+1$.
That means that we can set all $\alpha_j$'s equal as well.
Such choice automatically fulfils the selection rule, (\ref{selrule}).
Therefore the full four-point correlation function is given by
\begin{widetext}
\begin{eqnarray} \label{fulldisK}
{\cal K}_{+-}^{-+}(t_1,t_2; t_3, t_4) \approx Y(t_1,t_2; t_3, t_4)
\exp \Big( - \frac{e^2}{8 \pi f} \Big\{ \ln^2 \left|
\frac{f^4(t_1-t_4)(t_4-t_3)(t_3-t_2)(t_1-t_2)}{d^2
(t_1-t_3)(t_4-t_2)} \right| \Big\} \Big)
\end{eqnarray}
\end{widetext}
with
\begin{eqnarray} \nonumber
Y(t_1,t_2; t_3, t_4) \nonumber =  \left[ (\gamma^{1} \Lambda(t_1,t_4)
\gamma^{1})
\otimes (\gamma^{1} \Lambda(t_3,t_2) \gamma^{1}) \right]_* \, .
\end{eqnarray}
The latter expression is basically the non-interacting four-point
correlation function with special time ordering and is equal to
$G^{+-}(t_1-t_4) G^{+-}(t_3-t_2)$. The second term in
Eq.~(\ref{doubleW}) can be calculated in a similar manner
resulting in the same, up to an interchange of time variables,
exponential factor as in Eq.~(\ref{fulldisK}) and an appropriate
pre-factor. The pre-factors can in principle be re-exponentiated
but this would only results in logarithmic corrections which can
be neglected in comparison to the main disorder part. In order to
use formula (\ref{com}), we substitute $t_1=\tau_1$, $t_3=0$,
$t_2=t+\tau_1$ and $t_4=t+\tau_1+\tau_2$. We are interested in
the low-energy regime $\omega/V \ll 1$. Therefore the lower limit
for the time integrations can be set to $1/D$ ($D$ being the
conductance band-width) if it causes divergences and to zero
otherwise. Thus we obtain
\begin{eqnarray} \nonumber
n_>(\omega) &\sim& \int_0^\infty \, d\tau_1 \, \int_0^\infty \,
d\tau_2 \, e^{i \omega (\tau_1 + \tau_2)} \\ \nonumber
&\times& \exp \left\{ - \frac{e^2}{8 \pi f}
\ln^2 \left| \frac{f^4(\tau_1+\tau_2)}{d^2 D} \right| \right\} \, .
\end{eqnarray}
Estimating the $\tau$-integrals for small $\omega$ yields the
final result for the TED in vicinity of the Fermi edge:
\begin{eqnarray}                                 \label{disorderresult}
n_>(\omega) \sim \exp \left[ -\frac{e^2}{8 \pi f}
\ln^2 \left( \frac{d^2 D}{f^4\omega}\right) \right] \, .
\end{eqnarray}
Thus we find that the combined effect of the disorder and the
Coulomb interaction is to suppress the TED towards the Fermi
surface. This result can be understood in terms of the well known
Altshuler-Aronov-Lee anomaly \cite{aal} in the density of states
of the primary electrons. Indeed, one of the important results of
the KA's work \cite{kamenev} was to show that the negative
logarithmic correction to the density of states can be
exponentiated (and becomes a double-log due to the long-range
Coulomb forces). Our formula (\ref{disorderresult}) achieves a
similar exponentiation for the TED. As in \cite{kamenev}, this is
only valid above certain energy scale $\epsilon^*$, due to the
on-set of the fluctuation effects (corrections to the saddle-point
approximation). In MWNTs the situation is further complicated by
the dimensional cross-over effect; the corresponding energy scale
$\epsilon^*\simeq d/L^2$ ($L$ being the tube's circumference) was
estimated in \cite{eg2001}, see also \cite{kopietz2001} for
further discussion.

\section{Field emission from carbon nanotubes}      \label{FE}
\begin{figure}
\includegraphics[scale=0.25]{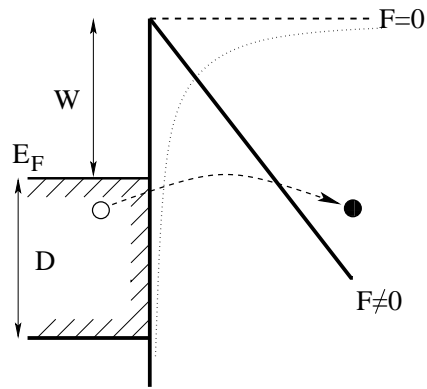}
\caption[]{\label{FEsetup} Schematic representation of the field
emission setup.  Here $F$ is the applied electric field, $W$ is
the work function and $D$ is the conduction band-width. The
dashed line represents the confining potential without field and
the dotted line stands for the image potential.}
\end{figure}

Field emission setup is relatively simple. An electric field of
strength $F$ is applied between a metallic tip of the emitter and
a counter-electrode which are usually placed in a vacuum.
Superposition of the confinement potential, the image potential
and the electrostatic potential due to the external electric
field leads to a formation of a barrier with a finite width
allowing the electrons to tunnel out of the emitter even at zero
temperatures, Fig.~\ref{FEsetup}.
Therefore the FE process can be regarded as tunnelling into
vacuum through a triangularly-shaped barrier whose upper part is
rounded off by the image potential. However, at low temperatures
the larger part of tunnelling electrons will have energies close
to the Fermi energy, so the exact form of the barrier is
unimportant.

Quantities of interest are the energy-resolved current
$j(\omega)$ and the total current $J$.
If the tunnelling amplitude is small (which it is for
all real systems), then $j(\omega)$ is proportional to the
probability $n(\omega)$ for an electron to have energy $\omega$,
to the energy dependent transmission coefficient $\cal D (\omega)$,
and to a factor $\cal F$ responsible for the tip geometry:
\begin{eqnarray} \label{fnappr}
j(\omega) = {\cal F} {\cal D}(\omega) n(\omega) \, .
\end{eqnarray}
In the noninteracting case and neglecting higher-order tunnelling
processes, $n(\omega)$ coincides with the Fermi distribution
function multiplied by the LDOS. The (quasi-classical) transmission
probability for a triangular barrier is given by \cite{landau}:
\begin{eqnarray}                          \label{tansmprob}
{\cal D}(\omega)\sim \exp( -4\sqrt{2m}(W-\omega)^{3/2}/3\hbar F ) \, ,
\end{eqnarray}
where $m$ is the electron mass and $W$ is the work function. As
discussed e.~g. in Ref.~\cite{plummer}, the transmission
coefficient is only slightly affected by the image potential and
for the field strength $F \ll (W-E_F)^2/e^3$ the barrier can be
regarded as strictly triangular. At zero temperatures, the
emerging (primary) spectrum has a sharp threshold at the Fermi
energy (no particles above $E_F$) and is essentially constant
in its vicinity.

The general picture does not change much even in the case of
interacting electrons. For the LLs $n(\omega)$ is proportional to
the relevant LDOS, which is known \cite{book,glazman} to be
\begin{eqnarray}\label{LLraspr}
n(\omega) = \Theta(-\omega) |\omega|^{1/g-1}/a_0 D^{1/g}
\Gamma(1/g) \, .
\end{eqnarray}
Plugging this into Eq.~(\ref{fnappr}) one observes that
at this lowest order in the tunnelling the TED above the Fermi energy
is still zero, even for the strongly correlated LL.
Below the Fermi energy the TED has a power-law singularity.
Expanding the transmission coefficient in powers of $\omega/W$ (work function
being the largest energy scale),
and integrating over all energies we establish the
Fowler-Nordheim (FN) formula for LLs:
\begin{eqnarray}\label{FN}
 J = \frac{\cal F}{a_0 D^{1/g}}
\left[ \frac{F^2}{4k_F W} \right]^{1/2g}
\exp\left( - \frac{4 k_F^{1/2} }{3F}W^{3/2} \right) \, ,
\end{eqnarray}
relating the full current to the electric field's strength
\cite{fowler,plummer}.

The generalisation of these results for SWNTs is rather
straightforward. Indeed, these systems are known to be described
as four-channel LLs. Three channels $\phi_{c-}$ (charge-flavour),
$\phi_{s+}$ (total spin), $\phi_{s-}$ (spin-flavour) are
non-interacting. The fourth channel $\phi_{c+}$ (total charge, or
the plasmon mode) possesses the LL parameter
$K=(1+4U_0/\pi)^{-1/2}$, where $U_0$ is the zero Fourier
component of the screened Coulomb potential. Note that though we
now have four channels the field-operator actually factorises as
\cite{sammlung1,sammlung2,sammlung3}
\begin{equation} \label{factor}
\psi\sim \exp\{i \phi_{c+}/(2\sqrt{K})+i(\phi_{c-}+
\phi_{s+}+\phi_{s-})/2\} \, .
\end{equation}
Just as in the case of the spinless LL, the interaction constant $K$
is included into the rescaled Bose fields and disappears from
the Hamiltonian, which is given by
\begin{eqnarray}
H=\frac{1}{4\pi}
\sum_{\delta=c,s j=\pm} \int \, dx \, (\partial_x \phi_{\delta j})^2 \, .
\end{eqnarray}
All the correlation functions also factorise. Hence the results
for the LDOS \cite{book,glazman} as well as the results of Section
\ref{luttinger}, Eqs.~(\ref{cresult}) and (\ref{exprelation}),
are still valid for SWNTs given that the substitution
\begin{equation}
g^{-1}\rightarrow (K^{-1}+3)/4 \, .
\label{subs}
\end{equation}
is made.

The typical value for $K$ in SWNTs lies near $0.2$ ($0.15$ to
$0.3$), which fixes the effective $g$ to $\approx 0.5$
\cite{sammlung1,sammlung2,sammlung3,tans1,bockrath1}. As a
result, Eq.~(\ref{FN}) is reasonably well approximated by the
classical FN law. This fact offers one possible explanation as to
why the experimental data of Refs.~\cite{french1,japan} can be
relatively well fitted by the conventional FN curve. In addition,
all the experiments currently available were made on SWNT films
where the nanotubes build a network with essentially a random 2D
geometry. This case, when extra complications are bound to arise,
is beyond the scope of this paper. To straightforwardly reveal
the correlation effects in the primary current, measurement on a
{\it single} SWNT are more appropriate. To our knowledge, this
has not been done yet.

Let us now discuss the secondary current. The analysis of the
secondary effects can also be made in terms of tunnelling into
vacuum. Thereby we should bear in mind that the theory presented
in Section \ref{general} assumes that the tunnelling amplitude is
energy independent. Therefore, since the effective applied
voltage is large in the FE setup (one has to send the chemical
potential on the right electrode  to minus infinity), some
integrals inevitably diverge. To overcome this difficulty and to
made the theory more quantitative, we now recall that there are
two candidates that can be used as the effective voltage. The
first one is the characteristic energy scale in the exponential
transmission coefficient ${\cal A} = F/2(k_F W)^{1/2}$, see
Eq.~(\ref{tansmprob}). The second one is the width of the emitter
conductance band $D$. Therefore the high-energy cutoff
playing the role of $V$ in the FE setup is either $D$ or ${\cal
A}$, whichever is smaller. We shall use $D$ through the rest of
the paper. With this modification all results of the previous
Sections for the TED $n(\omega)$ are valid with the understanding
that the tunnelling probability is proportional to the
transmission coefficient of the barrier at the Fermi energy,
$\gamma^2 \sim {\cal D}(E_F)$. According to Eqs.~(\ref{tokexp})
and (\ref{tokexp1}) the secondary current is proportional to the
TED. For SWNT the TED possesses all the singularities
characteristic for LLs, which we have already discussed. One
thing worth noting is that the critical value of the coupling now
is $K_c=1/5$, which is actually within the experimental range.
Therefore we do not make a specific prediction regarding the
character of the singularity (divergent versus convergent TED at
the Fermi edge). Instead we think that both the divergent and the
vanishing TEDs can be observed depending on the experimental
setup. Unfortunately, in most recent experiments, both
on SWNTs and MWNTs, only the total current is measured. Where the
energy-resolved measurements were actually made
\cite{french1,french2,japan}, in all cases but one, the
temperatures were too high for the secondary effects to be
visible. To our knowledge, so far only the experiments of
Ref.~\cite{fransen} contain high-energy tails which can be
attributed to secondary emission. However, the quality of the
presented data hinders us from attempting to actually fit the
curves.

\begin{figure}
\includegraphics[scale=0.3]{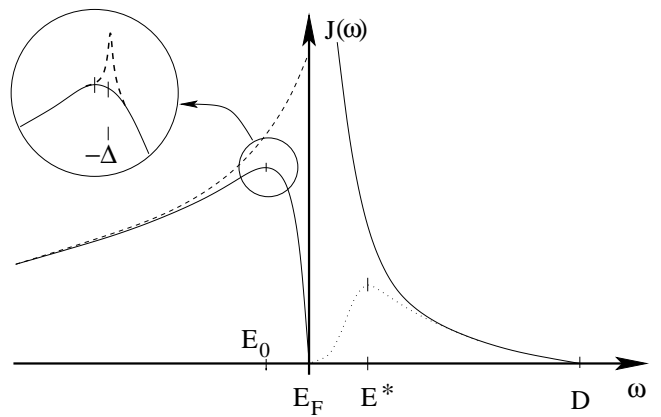}
\caption[]{\label{scenario} A sketch of the energy resolved current in a field
emission process from LLs. An additional Lorentz peak due to
tunnelling through a localised level (if exists) is depicted in the inset.}
\end{figure}

We conclude this Section by summarising our scenario for the
energy resolved current in the FE from LLs (and SWNTs). Well
below the Fermi energy $j(\omega)$ is governed by the exponential
growth of the transmission coefficient, $j(\omega) \sim
\exp(\omega/{\cal A})$, see Fig.~\ref{scenario}, as is the case
for any metallic emitter. Nearing the Fermi level, at energies
given by $E_{0} \approx (1-1/g) {\cal A}$, the non-linear LL LDOS
effects win over the exponential growth resulting in a maximum in
the current profile $j(\omega)$. Further towards the Fermi energy
it decreases according to the power-law: $\sim |\omega|^{1/g-1}$.
The edge behaviour right above $E_F$ is given by $\sim
\omega^{1/g-2}$. This changes shape depending on whether $g$ is
smaller or larger than the critical coupling $g_c$. In the case
of weak interactions, $1/2<g<1$, there is a singularity, while
for strong correlations, $0<g<1/2$ this singular behaviour is
suppressed and there is a power-law approach to zero. In the
latter case $j(\omega)$ acquires an additional maximum at a
cross-over energy $E^*$. For small applied fields, there is an
upper threshold at energy $\mbox{min}(D,{\cal A})$ where the
energy resolved current behaves as $\sim (\mbox{min}(D,{\cal
A})-\omega)^{1/g}$.

\section{Tunnelling via localised states} \label{LS}
Recent luminescence spectra measurements in FE experiments on
carbon nanotubes \cite{french1,fransen} suggest that, at least in some
cases, the emission process cannot be understood simply
in terms of tunnelling into vacuum.
They rather seem to be compatible with a model where
the emitted electron tunnels out of the system through one or more localised
states at the tip of the emitter \cite{french1,fransen}.
In this Section, we generalise our formalism in order
to analyse such a multiple stage tunnelling.

\begin{figure}
\includegraphics[scale=0.25]{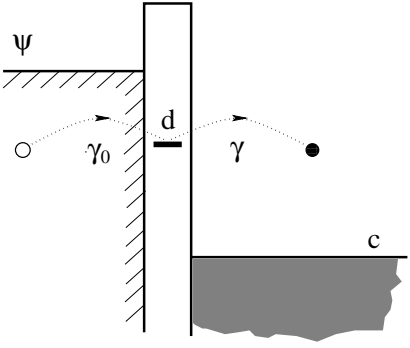}
\caption[]{\label{locur} A schematic representation of the
electron tunnelling through a localised state $d$.}
\end{figure}

We start with the Hamiltonian of the system.
Contrary to Eq.~(\ref{ham0}), the transfer of an electron between
the host and the lead occurs in two stages.
At the first stage, the electron populates a localised level with
energy $-\Delta$, see Fig.~\ref{locur}.
Let the tunnelling amplitude for this process be $\gamma_0$.
At the second stage the electron tunnels from the
localised state into the lead,
\begin{eqnarray}
 H_2 &=& H[\psi] + H_0[c] + \gamma_0 \left[ \psi(0)^\dag d + d^\dag \psi(0)
 \right] \nonumber \\
 &+& \gamma \left[ d^\dag c(0) + c^\dag(0) d \right] - \Delta d^\dag d \; ,
\end{eqnarray}
where we retain the notation $\gamma$ for the tunnelling amplitude
of the second process and $d^\dag,d$ are the creation and
annihilation operators of the localised state, respectively. We
assume that $0 < -\Delta < -V$. This is not restrictive but
reasonable. Indeed, the first inequality reflects the fact that
the electric field applied to the emitter causes an effective
lowering of energy of the states close to the tip. The second
inequality is a trivial one because if the energy of the state
$d$ lied below both Fermi levels it would have always been
populated. For simplicity, we
neglect any electrostatic interaction between the localised state
and the leads and the Hubbard term for $d$. Like in the case of
the simple tunnelling, the lead is assumed to be uncorrelated.

The problem of a localised state hybridised with a
non-interacting continuum is, of course, exactly solvable
\cite{mahan}. Therefore one way to proceed would be to eliminate
the $\gamma$ hopping term from the Hamiltonian exactly and then
apply perturbation theory in $\gamma_0$. However, in the FE setup
$\Delta$ is expected to be much smaller than the conductance band
width $D$. This implies that $\gamma_0 \gg \gamma$. Therefore we
take an alternative route and first calculate the TED of the
electrons on the level $d$ as a function of $g^{-+}(\omega)$ in
the host ignoring tunnelling to the lead. (We did the
alternative calculation as well and, at a given order of
perturbation theory, obtained exactly the same results.) To
proceed we define the Keldysh Green's function of the localised
level:
\begin{eqnarray}                             \label{Ddef}
 D(t) = -i \langle T_C [d(t) d^\dag(0) S_C ]\rangle \, ,
\end{eqnarray}
where the $S$-matrix includes only the tunnelling
between the host and the localised level,
\begin{eqnarray}                             \label{newSMat}
 S_C = T_C \exp\left( -i \gamma_0 \int_C \, dt \,
 \left[ \psi^\dag(0,t) d(t) + d^\dag(t) \psi(0,t)
 \right] \right) \; . \nonumber
\end{eqnarray}
Expanding the Green's function (\ref{Ddef}) and function $\langle
T_C [d(t) \psi^\dag(0,t') S_C ]\rangle$ in powers of $\gamma_0$
one obtains a set of coupled equations for them. Eliminating the
latter object one then obtains the following Dyson-type equation
for the Keldysh Green's function:
\begin{eqnarray} \label{dysonequation}
 D(t,t') = D_0(t,t') + \int_C \, d t'' \, D(t,t'') \kappa(t'',t') \; ,
\end{eqnarray}
where the kernel is given by
\begin{equation}
 \kappa(t'',t') =  \gamma_0^2 \int_C \, d t \, g_0(t'',t) D_0(t,t') \, .
\end{equation}
Here $g_0(t,t')$ and $D_0(t,t')$ denote the Green's functions of
the host and the localised level in the absence the hopping term.
Disentangling the Keldysh indices and Fourier transforming we
obtain the following set of equations:
\begin{eqnarray}
 D^{-+}(\omega) = D^{-+}_0(\omega) + D^{--}(\omega) \kappa^{-+}(\omega) -
 D^{-+}(\omega) \kappa^{++}(\omega) \; , \nonumber \\
 D^{--}(\omega) = D^{--}_0(\omega) + D^{--}(\omega) \kappa^{--}(\omega) -
 D^{-+}(\omega) \kappa^{+-}(\omega) \; . \nonumber
\end{eqnarray}
The solution is given by
\begin{widetext}
\begin{equation} \nonumber
 D^{-+}(\omega) = \frac{D^{-+}_0 + \gamma_0^2 g^{-+}_0(\omega)
 |D^{--}_0(\omega)|^2}{1+2 \gamma_0^2 \mbox{Re} [ g^{--}_0(\omega)
D^{--}_0(\omega)] + \gamma_0^4 |g^{--}_0(\omega) D^{--}_0(\omega)|^2} \, .
\end{equation}
\end{widetext}
This equation can be further simplified if one takes into account
that $\mbox{Re}[g^{--}_0(\omega)]/|D^{--}_0(\omega)|^2$  vanishes for
our system. Then the TED on the localised level is proportional to
\begin{equation}                     \label{correctD-+}
 D^{-+}(\omega) = \frac{\gamma_0^2 g^{-+}(\omega)}{(\omega+ \Delta)^2 +
 \gamma_0^4 |g^{--}(\omega)|^2} \, .
\end{equation}
In the last relation we omitted the subscript $0$ of $g(\omega)$
functions. Strictly speaking this constitutes an approximation.
The reason is that since the tunnelling onto the localised level
affects the Green's function of the host they have to be
calculated self-consistently. However, the renormalisations
occurring close to the Fermi energy should not strongly affect
the shape of the localised state and vice versa. That is unless
the localised level becomes resonant ($\Delta\to 0$), which case
we shall not consider in this paper. 
(We note though that the resonant level case can be approached
via a mapping onto the Kondo problem \cite{resonantlevel}.) 
For a non-interacting host
$|g^{--}(\omega)|^2$ is simply a constant equal to $1/4v_F^2$.

As we know from Section \ref{general}, the TED in the
non-interacting lead, is proportional to $D^{-+}(\omega)$ and the
tunnelling probability $\gamma^2$:
\begin{equation} \label{nf}
n(\omega) = -i \gamma^2 D^{-+}(\omega) \, .
\end{equation}

We now briefly discuss the picture for a non-interacting emitter.
According to Eq.~(\ref{correctD-+}) there is a Lorentzian peak at
$\omega=-\Delta$ with  height $4 v_F^2
g^{-+}(-\Delta)/\gamma_0^2$ and width $\gamma_0^2/2 v_F$. At the
second order in $\gamma^2$, there still is a sharp threshold at
the Fermi energy. These results for non-interacting emitters are,
of course, known (see e.~g.~\cite{plummer}). Here we have merely
re-derived them using the Keldysh formalism.

Let us now turn to the open question of how the presence of a
localised state would modify the FE process from a LL. For the
primary current we can still use formula (\ref{fnappr}), where we
have to substitute the TED of the electrons in the host by that
of the electrons on the localised state. The energy resolved
current is then given by the following expression (we again set
$v_F=1$),
\begin{eqnarray} \label{LSFNbare}
{\cal J}(\omega) &=& -i {\cal F} \, {\cal D}(\omega) \,
D^{-+} (\omega) \nonumber \\
&=& \frac{{\cal F} \, \gamma^2 \gamma_0^2 \, \Theta(-\omega)}{a_0
\, D^{1/g}\, \Gamma[1/g]} \,
\frac{|\omega|^{1/g-1}}{(\omega+\Delta)^2+\gamma_0^4/4} \nonumber \\
&\times& \exp\left( - 4 k_F^{1/2} \frac{(W-\omega)^{3/2}}{3F} \right) \, .
\end{eqnarray}
In comparison to the situation without the localised level, there
is an additional feature: the Lorentzian resonance
at $\omega= - \Delta$. To compute the total emitted current we
re-write the energy integral in terms of the dimensionless
variable $\xi = \omega/{\cal A}$ and expand the exponent in powers
of $\omega/W$ as the work function is large compared to other
energy scales. The total current then is
\begin{eqnarray}  \label{LSFNintegral}
{\cal J} &=& \frac{\gamma^2 \gamma_0^2 \, {\cal F}}{a_0 \, D^{1/g}
\, \Gamma[1/g]} \, \exp \left( - 4 k_F^{1/2}\, \frac{W^{3/2}}{3F}
\right) {\cal A}^{1/g-2} \nonumber \\&\times& \int_0^{\infty} \,
d\xi \, \frac{\xi^{1/g-1}}{(\xi- \Delta/{\cal A})^2 +
(\gamma_0^2/2 {\cal A})^2} \, e^{-\xi} \, .
\end{eqnarray}
This integral can be calculated exactly, resulting
in the modified FN relation
\begin{eqnarray} \nonumber
 {\cal J} = J \, \frac{2 \gamma_0^2}{\cal A} \,  \mbox{Im}  \,
 \left\{ w^{1/g-1} e^{w} \Gamma[1-1/g,w] \right\} \,
\end{eqnarray}
applicable for arbitrary parameters. Here $J$ is the total
current in the absence of the localised state, see
Eq.~(\ref{FN}), $\Gamma$ stands for the (incomplete) gamma
function and we have introduced a dimensionless quantity $w$,
defined by
\begin{eqnarray} \nonumber
 w =  \frac{1}{\cal A}\left( i \gamma_0^2/2 - \Delta \right) \, .
\end{eqnarray}

There two important limiting cases.

{\bf (i)} The electric field is weak, $\Delta/{\cal A} \gg 1$, or
$\Delta \gg F/2(k_F W)^{1/2}$. Then the resonance peak is far
away from the Fermi edge in comparison with the characteristic
decay scale of the transmission coefficient. The result of the
integration in Eq.~(\ref{LSFNintegral}) is then essentially
identical to the result for a system without a localised level
Eq.~(\ref{FN}), up to numerical pre-factors. Therefore we would
not expect that in this regime one can differentiate between the
direct tunnelling and tunnelling via a localised state.

{\bf (ii)} The electric field is sufficiently strong, so that
$\Delta/{\cal A} \ll 1$. Now the distance from the resonance to
the Fermi edge is much smaller than the characteristic energy
scale ${\cal A}$. If also $\gamma^2\ll\Delta$ (which is to be
expected) the total current is still given by Eq.~(\ref{FN}), apart from
the overall pre-factor, but with a different exponent: $1/2g$
should now be substituted by $1/2g-1$. This change can be
important for the analysis of experimental data on the SWNTs. For
SWNTs in this regime, we expect the modified FN plot to be linear in
the coordinates $\ln [{\cal J}/F^{(1/K-5)/4}]$ and $1/F$. For $K
\approx 0.2$ the total current depends exponentially on the field
strength (not a stretched exponential).

The calculation of the secondary current is more delicate.
In the case of the localised state the
insertion in the diagram for $g^{-+}(\omega)$,
see Figs.~\ref{fig3} and \ref{fig4}, should be modified.
It turns out that there is only one way to dress
the Green's function $G^{+-}_0$ with
the Green's functions of the localised level,
$G^{+-}_0(\omega) \rightarrow D^{++}(\omega) G^{+-}_0(\omega) D^{--}(\omega)$,
so that the diagram is non-zero above the Fermi edge.
The time- and anti-time-ordered localised level Green's
functions take the form
\begin{eqnarray} \nonumber
 D^{--(++)}(\omega) = \mp \frac{1}{\omega+\Delta
\pm i(\alpha_d+\gamma_0^2/2)} \, .
\end{eqnarray}
The imaginary part in the denominators reflects the fact that the
natural width of the localised level $\alpha_d$ is now widened by
the hybridisation with the continuum of states in the host. Using
the Fourier transform of the new insertion, which must now  be
substituted instead of the Fourier transform of
$G^{+-}_0(\omega)$ (given in Eq.~(\ref{com}) by
$\exp{i(\omega-V)t}/(t+i \alpha)$), we construct the localised
level counterpart of Eq.~(\ref{com}):
\begin{widetext}
\begin{eqnarray} \label{com2}
n_>(\omega) &=& i \frac{\gamma^2}{2 \pi (\alpha_d + \gamma_0^2)}
\int_{-\infty}^\infty  dt e^{i (\omega+\Delta)t} (e^{(\alpha_d +
\gamma_0^2)t} \Gamma[t(\alpha_d + \gamma_0^2 + i (\Delta+D))] -
e^{-(\alpha_d + \gamma_0^2)t} \Gamma[t(-\alpha_d - \gamma_0^2 + i
(\Delta+D))]) \nonumber \\ &\times& \int_0^\infty d\tau_1
\int_0^\infty d\tau_2e^{i\omega(\tau_1+\tau_2)} {\cal
K}(\tau_1,0;t+\tau_1,t+\tau_1+\tau_2) \, .
\end{eqnarray}
\end{widetext}
The asymptotic behaviour of
the $n_>(\omega)$ close the Fermi edge is not affected (apart from a
numerical pre-factor) by the presence of the localised state.
There is an important difference -- the upper threshold for the
TED is now given by $\Delta$ instead of $D$. The reason is that in
the presence of the localised state the upper energy limit of the
hot hole is given by the energy of the $d$-level instead of the
band-width $D$ (or the applied voltage $V$).

Thus, in the case of tunnelling via a localised state the TED
obtains an additional feature, namely a Lorentzian peak below the
Fermi edge, see the inset in Fig.~\ref{scenario}. Recent
luminescence experiments of Ref.~\cite{french1} suggest a similar
structure -- two superimposed peaks in the spectrum of emitted
light. The Lorentzian shape of the peaks was probably hidden by
the intrinsic Gaussian broadening introduced by the apparatus (a
Gaussian fit was adopted in Ref.~\cite{french1}). The calculation
of the emerging luminescence spectra is beyond the scope of this
paper and will be discussed elsewhere.

\section{Summary and conclusions} \label{conclusions}
In this paper,  we studied the energy resolved current (or the
TED) emitted from a tip of a correlated host material under
out-of-equilibrium conditions. In the uncorrelated case the TED
has a sharp threshold at the Fermi edge and there are no
particles with energies above it. In the presence of interactions
there is a finite high-energy tail due to the hot hole relaxation
process. For most systems, the TED is singular right above the
Fermi edge.

In the repulsive Luttinger liquid model, where the LDOS is
renormalised by interactions and vanishes towards the Fermi edge,
the TED below the Fermi edge shows a power-law behaviour with the
LDOS exponent. Above the Fermi energy, we also find a power-law
but with a new exponent. There is a critical coupling and the TED
is divergent for weak interactions but convergent for strong
interactions. In the latter case, the TED has a maximum above the
Fermi level. Although the high-energy tail still exists in
systems with local interactions confined to the tip, the TED is
then a regular function of the energy. Additional features in the
TEDs arise in a situation when the tunnelling occurs in two
stages via a localised state. In this case there is a Lorentzian
peak centred at the energy of this localised level and of the
width which depends on the hybridisation with the continuum
states on both sides of the contact. All the singularities of the
TED above the Fermi energy survive the introduction of the
localised level. The singularities in high-energy tails are
suppressed as soon as the disorder is introduced. We have
calculated the TED of particles tunnelling from a correlated
disordered 2D system. In this case the LDOS is exponentially
suppressed and this suppression is carried over to the TED.
Contrary, in the pure limit, we find a weak (logarithmic)
singularity in the TED above the Fermi edge.

We then specialised our results to the case of the FE from
carbon nanotubes, both SWNTs and MWNTs.
While they are in qualitative agreement with existing experiments,
there is little data currently available to test our predictions
for the secondary current in these systems.
We hope that this work may encourage more measurements,
especially at low temperatures ($\ll$ 300K).
This could provide additional information about the nature of
interactions in nanotubes.

Throughout the paper we consistently worked at zero
temperatures. All our results can be generalised to
finite temperatures by standard methods (though
this may not be of immediate interests until there
is more experimental data).
There are, of course, other open questions (like the role
of higher orders in the tunnelling amplitude) and potentially
interesting extensions of this work, which are detailed
in the main text.

Also interesting, though more difficult, could be experiments on
tunnelling junctions if the TED in such setups were accessible
e.~g. by means of a tunnel probe \cite{devoret}. In this case both
threshold asymptotics could reveal essential details about the
correlations. Even if direct measurements of the TED are not
possible, there is another possibility to identify the secondary
effects. Since the hot hole decay is a relaxation process, an
electron-hole recombination can occur with an irradiation of a
photon. Such a current-driven luminescence effect could be
observable in quantum dot systems, where the corresponding
experimental techniques became widespread in the recent years
\cite{dekel}.

\acknowledgments We have benefited from illuminating discussions
with Nathan Andrei, David Edwards, Reinhold Egger, Fabian Essler,
and Yang Chen. This work was supported by the EPSRC of the UK
under grants GR/N19359 and GR/R70309. The authors also participate
in the EC training network DIENOW.

\bibliography{prb}

\end{document}